\documentclass[10pt, conference, letterpaper]{IEEEtran}
\IEEEoverridecommandlockouts

\newif\ifhideappendix

\hideappendixfalse 

\newcommand{\apxonly}[1]{\ifhideappendix\else#1\fi}

\usepackage[T1]{fontenc}
\usepackage[utf8]{inputenc}
\usepackage{cite}
\usepackage{amsmath,amssymb,amsfonts}
\usepackage{graphicx}
\usepackage{textcomp}
\usepackage{xcolor}
\usepackage{url}
\usepackage{lipsum}

\usepackage[nolist,nohyperlinks]{acronym}

\begin{acronym}
  \acro{AP}{access point}
  \acro{ACK}{acknowledgment}
  \acro{BA}{block acknowledgment}
  \acro{BAR}{block acknowledgment request}
  \acro{MU}{multi user}
  \acro{MU-BAR}{multi-user BA request}
  \acro{CTMC}{continuous time Markov chain}
  \acro{LLM}{large language model}
  \acro{KL}{Kullback-Leibler}
  \acro{A-MPDU}{aggregated MAC protocol data unit}
  \acro{STA}{station}
  \acro{MAPC}{multi-AP coordination}
  \acro{BSS}{basic service set}
  \acro{SINR}{signal-to-interference-plus-noise ratio}
  \acro{RSSI}{received signal strength indicator}
  \acro{RL}{reinforcement learning}
  \acro{AI}{artificial intelligence}
  \acro{ML}{machine learning}
  \acro{DRL}{deep reinforcement learning}
  \acro{MA-DRL}{multi-agent deep reinforcement learning}
  \acro{GNN}{graph neural network}
  \acro{FM}{flow matching}
  \acro{MDN}{mixture density network}
  \acro{ODE}{ordinary differential equation}
  \acro{EMA}{exponential moving average}
  \acro{AE}{autoencoder}
  \acro{VAE}{variational autoencoder}
  \acro{MPNN}{message-passing neural network}
  \acro{MLP}{multi-layer perceptron}
  \acro{GELU}{Gaussian error linear unit}
  \acro{PCA}{principal component analysis}
  \acro{MMD}{maximum mean discrepancy}
  \acro{NLL}{negative log-likelihood}
  \acro{MSE}{mean squared error}
  \acro{MCS}{modulation and coding scheme}
  \acro{Co-SR}{coordinated spatial reuse}
  \acro{Co-TDMA}{coordinated time division multiple access}
  \acro{Co-BF}{coordinated beamforming}
  \acro{TXOP}{transmission opportunity}
  \acro{MAB}{multi-armed bandit}
  \acro{H-MAB}{hierarchical multi-armed bandit}
  \acro{UCB}{upper confidence bound}
  \acro{ELBO}{evidence lower bound}
  \acro{CCA}{clear channel assessment}
  \acro{MCMC}{Markov chain Monte Carlo}
  \acro{PRS}{pure random search}
  \acro{GDM}{generative diffusion model}
  \acro{GAN}{generative adversarial network}
  \acro{CDF}{cumulative distribution function}
  \acro{DCF}{distributed coordination function}
  \acro{GPU}{graphics processing unit}
  \acro{RWPM}{random waypoint mobility model}
\end{acronym}

\usepackage{pgfplots}
\usepgfplotslibrary{statistics}
\pgfplotsset{compat=1.18, every axis plot/.append style={thick}}

\usepackage{tikz}
\usetikzlibrary{arrows.meta,positioning,calc,shapes.geometric,fit}

\usetikzlibrary{decorations}
\usetikzlibrary{decorations.pathmorphing}
\usetikzlibrary{decorations.pathreplacing}
\usetikzlibrary{shapes}
\usetikzlibrary{through}
\usetikzlibrary{backgrounds,positioning}

\definecolor{green}{RGB}{0,176,80}
\definecolor{blue}{RGB}{15,158,213}
\definecolor{red}{RGB}{255,0,0}

\definecolor{palA}{RGB}{58,82,139}
\definecolor{palB}{RGB}{39,124,142}
\definecolor{palB2}{RGB}{33,145,140}
\definecolor{palC}{RGB}{31,163,134}
\definecolor{palD}{RGB}{94,201,97}

\definecolor{palE}{RGB}{204,71,20}
\definecolor{palF}{RGB}{225,100,98}
\definecolor{palG}{RGB}{242,132,75}
\definecolor{palH}{RGB}{252,166,54}

\usepackage{subfig}

\usepackage{booktabs}
\usepackage{multirow}
\usepackage{siunitx}
\begin{document}

\title{FM4WiFi: Flow Matching for Multi-AP Coordination in Dense Deployments of Beyond Wi-Fi 8 Networks
\ifhideappendix
\else
    \thanks{This research was funded by the National Science Centre, Poland (2023/05/Y/ST7/00004). We gratefully acknowledge Polish high-performance computing infrastructure PLGrid (HPC Center: ACK Cyfronet AGH) for providing computer facilities and support within computational grant no. PLG/2025/018718. We gratefully acknowledge the support of the National Research Institute, grant number POIR.04.02.00-00-D008/20-01 on ``National Laboratory for Advanced 5G Research'' (acronym PL-5G) for providing computing facilities.
    
    For the purpose of Open Access, the authors have applied a CC-BY public copyright licence to any Author Accepted Manuscript (AAM) version arising from this submission.}
\fi
}

\ifhideappendix
    \author{\IEEEauthorblockN{Anonymous Author(s)}}
\else
    \author{
    \IEEEauthorblockN{Maksymilian Wojnar,
    Krzysztof Rusek,
    Katarzyna Kosek-Szott, and
    Szymon Szott}
    \IEEEauthorblockA{
    AGH University of Krakow, Poland\\
    Email: \{name.surname\}@agh.edu.pl}
    }
\fi

\maketitle

\begin{abstract}
Wi-Fi networks are moving beyond random channel access toward tightly coordinated operation across access points (APs), a shift reflected in Wi-Fi~8's multi-AP coordination (MAPC). However, the current  MAPC specification restricts cooperation to AP pairs, fundamentally limiting the gains achievable in dense deployments and calling for scalable, network-wide coordination in beyond Wi-Fi~8 systems.
We target coordinated spatial reuse (Co-SR), where APs transmit concurrently at reduced power. Effective Co-SR demands joint selection and configuration of AP--station transmissions, yet existing approaches simply do not scale: they rely on heavy signaling, slow convergence, unrealistic assumptions, and often require computation time that explodes with network size.
We introduce FM4WiFi, a generative ML pipeline that addresses these limitations by producing high-quality Co-SR configurations in a single inference step. FM4WiFi integrates (i) an autoencoder that learns compact latent representations of network states, (ii) a flow-matching generative model that synthesizes feasible Co-SR configurations (including rate control, absent from prior work), and (iii) a surrogate rate predictor that allows rapid, large-scale Co-SR candidate evaluation without dependence on a live system or digital twin.
Across extensive evaluations (including experimental validation), FM4WiFi matches or exceeds state-of-the-art baselines at medium-to-large scales and scales to 30+~APs with sub-second inference.
\apxonly{Extensive ablation studies validate each modeling and optimization choice.}
\end{abstract}


\section{Introduction}

Wi-Fi networks are undergoing a shift from competitive, randomly distributed channel access to collaborative coordination among \acp{AP}. 
This transition is reflected in one of the defining features of the forthcoming Wi-Fi 8 standard: \ac{MAPC}. 
However, MAPC's design limits cooperation to pairs of APs, restricting its potential benefits. 
Consequently, we expect that future beyond Wi-Fi~8 systems will need to support scalable, network-wide coordination strategies.

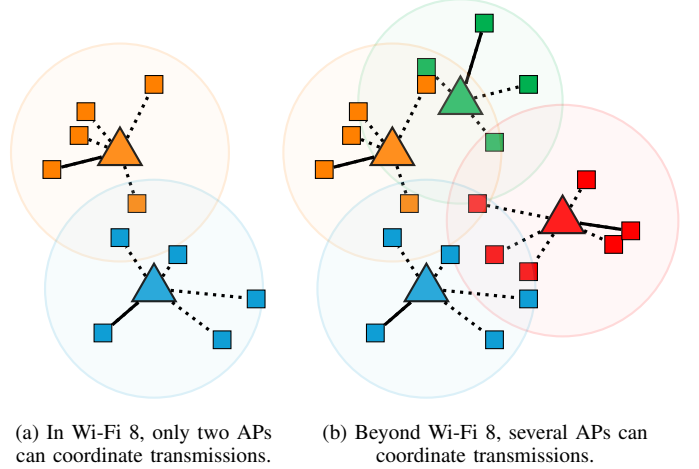
\begin{figure}[!t]
\resizebox{\columnwidth}{!}{%
\begin{tikzpicture}
\def\panelgap{3.6}   

\begin{scope}[scale=0.45]

\node[regular polygon, regular polygon sides=3, draw, thick, fill=orange]
    (Lap4) at (-2,-1.5) {};
\draw[orange, thick, opacity=0.15, fill=orange!20] (-2,-1.5) circle (3.2cm);

\foreach \i/\x/\y in {1/-3/-0.3, 2/-1/0.5, 3/-4/-2.0, 4/-1.5/-3.0, 5/-3.2/-1.0} {
    \node[rectangle, draw, fill=orange] (LstaD\i) at (\x,\y) {};
    \draw[dotted, line width=1.2] (Lap4) -- (LstaD\i);
}
\draw[line width=1.2] (Lap4) -- (LstaD3);

\node[regular polygon, regular polygon sides=3, draw, thick, fill=blue]
    (Lap6) at (-1,-5.5) {};
\draw[blue, thick, opacity=0.15, fill=blue!20] (-1,-5.5) circle (3.2cm);

\foreach \i/\x/\y in {1/-2.5/-6.8, 2/1/-7.0, 3/-0.3/-4.5, 4/2/-5.8, 5/-2/-4.0} {
    \node[rectangle, draw, fill=blue] (LstaF\i) at (\x,\y) {};
    \draw[dotted, line width=1.2] (Lap6) -- (LstaF\i);
}
\draw[line width=1.2] (Lap6) -- (LstaF1);

\end{scope}

\begin{scope}[shift={(\panelgap cm,0)}, scale=0.45]

\node[regular polygon, regular polygon sides=3, draw, thick, fill=green]
    (ap1) at (0,0) {};
\draw[green, thick, opacity=0.15, fill=green!20] (0,0) circle (3.0cm);

\foreach \i/\x/\y in {1/1/-1.2, 2/-1/1, 3/2/0.5, 4/0.7/2.3} {
    \node[rectangle, draw, fill=green] (staA\i) at (\x,\y) {};
    \draw[dotted, line width=1.2] (ap1) -- (staA\i);
}
\draw[line width=1.2] (ap1) -- (staA4);


\node[regular polygon, regular polygon sides=3, draw, thick, fill=red]
    (ap3) at (3,-3.5) {};
\draw[red, thick, opacity=0.15, fill=red!20] (3,-3.5) circle (3.4cm);

\foreach \i/\x/\y in {1/1/-4.5, 2/4.5/-4.2, 3/2/-5.0, 4/5/-3.8, 5/3.7/-2.3, 6/0.5/-3.0} {
    \node[rectangle, draw, fill=red] (staC\i) at (\x,\y) {};
    \draw[dotted, line width=1.2] (ap3) -- (staC\i);
}
\draw[line width=1.2] (ap3) -- (staC4);

\node[regular polygon, regular polygon sides=3, draw, thick, fill=orange]
    (ap4) at (-2,-1.5) {};
\draw[orange, thick, opacity=0.15, fill=orange!20] (-2,-1.5) circle (3.2cm);

\foreach \i/\x/\y in {1/-3/-0.3, 2/-1/0.5, 3/-4/-2.0, 4/-1.5/-3.0, 5/-3.2/-1.0} {
    \node[rectangle, draw, fill=orange] (staD\i) at (\x,\y) {};
    \draw[dotted, line width=1.2] (ap4) -- (staD\i);
}
\draw[line width=1.2] (ap4) -- (staD3);

\node[regular polygon, regular polygon sides=3, draw, thick, fill=blue]
    (ap6) at (-1,-5.5) {};
\draw[blue, thick, opacity=0.15, fill=blue!20] (-1,-5.5) circle (3.2cm);

\foreach \i/\x/\y in {1/-2.5/-6.8, 2/1/-7.0, 3/-0.3/-4.5, 4/2/-5.8, 5/-2/-4.0} {
    \node[rectangle, draw, fill=blue] (staF\i) at (\x,\y) {};
    \draw[dotted, line width=1.2] (ap6) -- (staF\i);
}
\draw[line width=1.2] (ap6) -- (staF1);

\end{scope}
\end{tikzpicture}%
}

\par\vspace{3pt}
\begin{minipage}[t]{0.40\columnwidth}\centering\footnotesize
(a) In Wi-Fi~8, only two APs can coordinate transmissions.
\end{minipage}\hfill
\begin{minipage}[t]{0.58\columnwidth}\centering\footnotesize
(b) Beyond Wi-Fi~8, several APs can coordinate transmissions.
\end{minipage}%
\caption{Example topologies and Co-SR scheduling in (a) Wi-Fi~8 and (b) beyond Wi-Fi~8 networks. Triangles denote APs, squares -- stations, lines -- viable concurrent transmissions, dotted lines -- associations.}
\label{fig:mapc_csr_example}
\end{figure}

We focus on \ac{Co-SR}, an MAPC variant, which is particularly promising for high-density deployments, e.g., industrial warehouses or stadiums.
\ac{Co-SR} scheduling requires selecting (i) \ac{AP} groups that transmit simultaneously, (ii) \acp{STA} these \acp{AP} transmit to, and (iii) Wi-Fi parameters (such as \ac{MCS} indices and transmit power levels) \cite{ning2025survey,val2025wi}. We refer to such a complete assignment as a \emph{\ac{Co-SR} configuration}.
Fig.~\ref{fig:mapc_csr_example} provides example topologies for \ac{Co-SR} in both Wi-Fi~8 and beyond Wi-Fi~8 networks. The scheduling problem in the latter case is compounded by the explosion in the number of possible configurations. 
Existing Co-SR scheduling solutions are not ready to operate on a network scale due to their signaling overhead, slow convergence times, computational complexity, and idealistic assumptions \cite{wojnar2025csr}.
Indeed, the four APs shown in Fig.~\ref{fig:mapc_csr_example} are at the limit for currently available scheduling solutions. 

\begin{figure*}[t]
    \centering
    \includegraphics[width=0.7\linewidth]{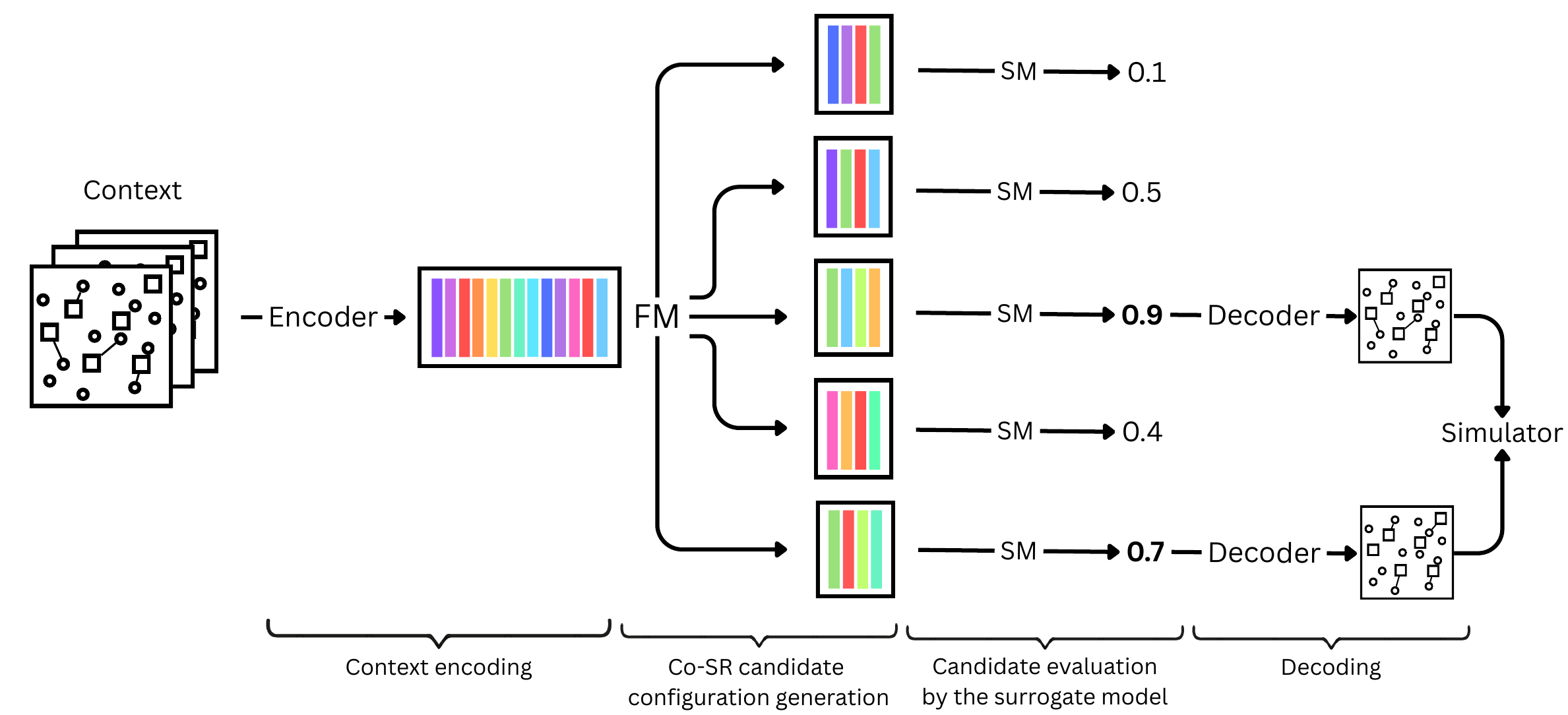}
    \caption{Full FM4WiFi inference pipeline. The central controller encodes recent network observations with the encoder, generates candidate \ac{Co-SR} configurations with the \ac{FM} model, scores them using the surrogate predictor, and decodes the top-$k$ candidates into scheduling decisions.}
    \label{fig:inference_pipeline}
\end{figure*}

\apxonly{
\begin{figure}[t!]
    \centering
    \includegraphics[width=\linewidth]{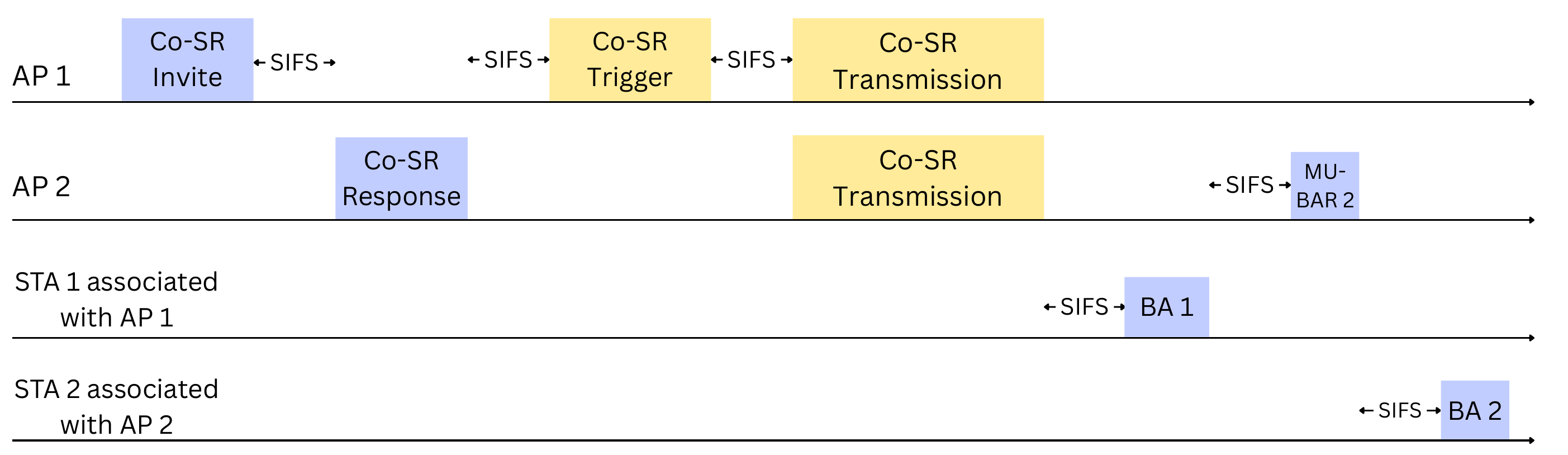}
    \caption{Co-SR frame exchange sequence \cite{802.11-26/0287r1}.}
    \label{fig:co-sr-frame-exchange}
\end{figure}
}

To address this challenge, we propose FM4WiFi: a generative pipeline that learns offline from a diverse dataset of network topologies and subsequently generates high-performance Co-SR transmission configurations at inference time. 
Our pipeline incorporates the latest advancements in \ac{ML}: \iac{GNN} autoencoder to learn latent representations of network states, a flow matching (FM) generative model to produce plausible Co-SR configurations, and a surrogate data rate predictor that enables rapid evaluation of Co-SR configuration candidates without the need to access a real network or its digital twin (Fig.~\ref{fig:inference_pipeline}).
During inference, the central controller collects data (\ac{RSSI} values and transmission outcomes) from all managed access points through wired connections, both regarding themselves and their associated stations. This allows the controller to understand the current state of the network (context), almost immediately select promising candidates for Co-SR transmissions, and apply the configuration to the Co-SR AP group, as elaborated in Section~\ref{sec:inference}.

\apxonly{
To clarify how \ac{Co-SR} operates at the MAC-layer, Fig.~\ref{fig:co-sr-frame-exchange} presents the Wi-Fi~8 Co-SR frame-exchange sequence. In this example, two APs perform simultaneous downlink Co-SR transmissions following an initiation exchange. The associated stations then acknowledge both transmissions using \ac{BA} frames. Additionally, AP~1 configures the \ac{ACK} policy as part of the Co-SR transmission; as a result, the second station's \ac{ACK} is explicitly requested by its AP via a \ac{MU-BAR} frame to avoid potential collisions.
This MAC-layer exchange illustrates the practical complexity of configuring Co-SR transmissions, which motivates the need for automated methods capable of optimizing such operations in larger, beyond-Wi-Fi~8 networks.
}

We extensively evaluate FM4WiFi across a wide range of scenarios (including experimental validation), demonstrating improvements over state-of-the-art baselines in data rate with unmatched scalability\apxonly{, and validating key design choices through ablation studies}.
Our contributions are as follows:
\begin{itemize}
    \item \textbf{FM4WiFi.} We propose the first generative pipeline in the literature for \ac{Co-SR} scheduling, built on a \ac{GNN} autoencoder with a novel mixture categorical loss for heterogeneous Wi-Fi attributes, a flow matching model for configuration generation, and a surrogate predictor for fast candidate ranking.
    \item \textbf{Joint scheduling and rate selection.} FM4WiFi joint\-ly optimizes \ac{Co-SR} group formation, station selection, \ac{MCS}, and transmit power in a single model, eliminating the idealized \ac{MCS} assumptions of all prior work.
    \item \textbf{Scalability and speed.} FM4WiFi produces schedules in under \SI{1}{\second} and scales to 30+~\acp{AP}, where the existing analytical and \ac{RL} Co-SR baselines become intractable.
    \item \textbf{Dynamic robustness.} Event-triggered regeneration enables immediate adaptation to topology changes, station joins, reassociations, and mobility without retraining in contrast to fixed-action-space \ac{RL} methods.
    \item \textbf{Full reproducibility.} The FM4WiFi pipeline and all code used for dataset generation and model training are available as open source.\footnote{\url{https://github.com/ml4wifi-devs/fm4wifi}}
    \item \textbf{Experimental testbed.} We provide the first open-source experimental testbed on commercial-off-the-shelf devices for evaluating Co-SR.\footnote{\url{https://github.com/ml4wifi-devs/mapc-testbed}}
\end{itemize}

\section{Motivation}
\label{sec:motivation}

Our work on developing a \ac{Co-SR} scheduler able to operate in dense Wi-Fi~8 networks is motivated by the following limitations of existing works.

\textbf{Vulnerability of RSSI-based centralized schemes.} 
Several centralized \ac{Co-SR} schemes choose \ac{AP} groups, their transmit power levels, and \ac{MCS} indices from global \ac{RSSI} or \ac{SINR} information available at a central controller~\cite{haxhibeqiri2024coordinated,nunez2023group}.
Although effective in small-scale scenarios, these architectures add coordination overhead as the network grows and depend on heuristic, parameter-tuned group formation. In~\cite{haxhibeqiri2024coordinated}, the stations wirelessly report the environment data to the controller, introducing signaling overhead, and \ac{MCS} selection relies on topology-dependent thresholds. Similarly,~\cite{nunez2023group} uses two key parameters that govern successful grouping, yet their selection is deferred to future work. In dense or mobile environments, \ac{RSSI} variance can perturb \ac{SINR} estimates, causing poor \ac{Co-SR} configuration.

\textbf{Scalability and convergence in probing-based methods.} 
To reduce dependence on unreliable measurements, online learning methods have been introduced~\cite{wojnar2025ieee, wojnar2025csr}. These methods probe different configurations to learn agent policies (i.e., mapping states to actions) without \textit{a priori} knowledge of the wireless channel. However, the convergence time still increases with the topology size and \ac{TXOP} duration, and large deployments typically require additional clustering to keep the action space manageable~\cite{wojnar2025csr}.

\textbf{Rigidity of traditional RL}
Recent discrete-action \ac{RL} schedulers~\cite{jung2026coordinated, nunez2025deep} select one action from a precomputed catalog of feasible \ac{Co-SR} groups, i.e., a fixed discrete output space with masking over topology-dependent options. This design does not generate new \ac{Co-SR} configurations online; it only schedules among configurations prepared in advance. Notably,~\cite{nunez2025deep} is specifically concerned with the temporal scheduling of pre-formed configurations to reduce latency, which is a complementary problem to configuration generation and not a direct competitor to our approach. Thus, adaptation to unseen deployments typically requires regenerating the action space and retraining, while convergence may require millions of interaction steps \cite{nunez2025deep}. In contrast, our approach can accept any topology at both input and output, making adaptation to unseen deployments an inherent property of the architecture rather than a separate engineering effort.

\textbf{Model-based optimization restrictions.}
Model-based approach\-es fall into two categories: solver-based and optimization-guided \ac{RL}. \textit{Solver-based formulations} can provide exact (or upper-bound) configurations~\cite{wojnar2025csr, zhu2025two}, but require accurate inputs (e.g., path loss or \ac{RSSI} measurements) and iterative solving, which is difficult to recompute at strict per-TXOP timescales in fast-varying settings. Meanwhile, \textit{optimization-guided \ac{RL} formulations}, such as~\cite{jung2026coordinated}, replace direct combinatorial search with \ac{MA-DRL}, but still rely on periodic channel-state updates and introduce substantial training complexity.

\textbf{Lack of joint MCS selection.} 
A critical but frequently simplified dimension of \ac{Co-SR} scheduling is the selection of appropriate \ac{MCS}. Many methods decouple this step or use a fixed \ac{MCS} to maintain model simplicity or tractability~\cite{jung2026coordinated, wojnar2025ieee, wojnar2025csr, zhu2025two}. This simplification fails to capture the interplay between interference management and feasible data rates, often leading to higher packet error rates or the reliance on unrealistic ideal \ac{MCS} selection.

\textbf{Generative approach.} 
We identify a common limitation across state-of-the-art approaches: they treat \ac{Co-SR} scheduling as either a \textit{search} over a fixed set of groups or a \textit{discrete classification} problem. This limits their ability to both generalize to arbitrary, non-grid topologies and provide fast and accurate joint optimization. In contrast, we treat \ac{Co-SR} group formation as a \textit{conditional generation} task. As summarized in Table~\ref{tab:comparison}, our framework is designed to provide the signaling efficiency, flexibility, and generalization needed for next-generation Wi-Fi coordination schemes.

\begin{table}[t]
  \caption{Comparison of \ac{Co-SR} scheduling approaches.}
  \label{tab:comparison}
  \centering
  \small
  \setlength{\tabcolsep}{4pt}
  \begin{tabular}{@{}lc ccccc@{}}
    \toprule
    Approach & Ref.
      & \rotatebox[origin=l]{90}{Scalable}
      & \rotatebox[origin=l]{90}{Low  overhead} 
      & \rotatebox[origin=l]{90}{Real-time}
      & \rotatebox[origin=l]{90}{Generalization}
      & \rotatebox[origin=l]{90}{Joint MCS opt.} \\
    \midrule
    \ac{RSSI}-based & \cite{haxhibeqiri2024coordinated,nunez2023group} & \texttimes & \texttimes & \checkmark & \checkmark & \checkmark \\
    Probing-based & \cite{wojnar2025ieee,wojnar2025csr} & \texttimes & \checkmark & \checkmark & \checkmark & \texttimes \\
    Fixed action-space \ac{RL} & \cite{nunez2025deep} & \texttimes & \checkmark & \checkmark & \texttimes & \texttimes \\
    Optimization (solver-based) & \cite{wojnar2025csr,zhu2025two} & \texttimes & \texttimes & \texttimes & \checkmark & \checkmark/\texttimes \\
    Optimization (\ac{RL}-assisted) & \cite{jung2026coordinated} & \texttimes & \texttimes & \checkmark & \texttimes & \texttimes \\
    \specialrule{0.4pt}{2pt}{2pt}
    \textbf{Generative model (ours)} & This paper & \checkmark & \checkmark & \checkmark & \checkmark & \checkmark \\
    \bottomrule
  \end{tabular}
\end{table}

\section{System Design}
\label{sec:system_design}

We propose a generative pipeline that learns offline from a diverse dataset of network topologies and subsequently generates high-performance \ac{Co-SR} configurations at inference time. The pipeline consists of three components: (i)~an autoencoder that learns latent representations of Wi-Fi network states, (ii) an \ac{FM} generative model that produces plausible \ac{Co-SR} configurations conditioned on recent network history, and (iii)~a surrogate data rate predictor that enables rapid evaluation of \ac{Co-SR} configuration candidates without the need for time-consuming simulation. The design of these components is grounded in a Bayesian view of the coordination problem, which we formalize next.

\subsection{Inductive Bias}
\label{sec:inductive_bias}

\ac{Co-SR} scheduling can be viewed as a joint process of inference and optimization. A coordination agent must deploy a configuration $c$ to maximize observed network throughput $T$. If the agent possessed complete environmental knowledge, i.e., exact station positions and propagation conditions, collectively denoted $\Theta$, this would reduce to a deterministic optimization problem~\cite{wojnar2025csr}. In practice, $\Theta$ is latent; only empirical observations $\mathcal{D}$ (packet outcomes, \ac{RSSI}) are available.

We formalize this as a Bayesian problem. The agent maintains a prior $p(\Theta)$ and uses $\mathcal{D}$ to form a posterior $p(\Theta|\mathcal{D})$. The optimal configuration is then:
\begin{equation}
    c^* = \arg\max_{c}\; \mathbb{E}_{\Theta \sim p(\Theta|\mathcal{D})}\bigl[T(c, \Theta)\bigr].
\end{equation}
Both the inference and optimization steps are analytically intractable for real-world deployments, motivating learned approximations.

The \ac{FM} model serves as the inference step: drawing diverse candidate configurations conditioned on recent observations in a single forward pass. This mirrors the role of \ac{MCMC} in Bayesian statistics, where a Markov chain converges to the posterior as its stationary distribution; \ac{FM} provides a tractable surrogate for this process. The surrogate data-rate predictor serves as the optimization step: acting as a learned digital twin of the network, it scores each candidate and replaces costly real-network evaluation, approximating the expectation under the posterior.

\subsection{Graph Representation of Wi-Fi Networks}
\label{sec:graph_repr}

\begin{figure}[t]
\centering
\scalebox{1.0}{
\begin{tikzpicture}[
    scale=0.62,
    apnode/.style={regular polygon, regular polygon sides=3, draw, thick,
                   minimum size=10pt, inner sep=0pt},
    stanode/.style={rectangle, draw, minimum size=6pt, inner sep=0pt},
    apedge/.style={green!60!black, dash dot, line width=1.2},
    staedge/.style={black, line width=1.2},
    staedgedotted/.style={black, dotted, line width=1.2},
    attrbox/.style={draw=black!60, fill=white, rounded corners=2pt,
                    font=\scriptsize, inner sep=3pt, align=left},
]

\node[apnode, fill=green]  (ap1) at (-1.3, 1.7) {};
\node[apnode, fill=blue]   (ap2) at (1.5, 2.2) {};
\node[apnode, fill=red]    (ap3) at (3.2, 0.2) {};
\node[apnode, fill=orange] (ap4) at (0.1, -0.7) {};

\draw[green,  thick, fill=green!20,  opacity=0.15]  (ap1) circle (1.9cm);
\draw[blue,   thick, fill=blue!20,   opacity=0.15]  (ap2) circle (1.8cm);
\draw[red,    thick, fill=red!20,    opacity=0.15]  (ap3) circle (2.0cm);
\draw[orange, thick, fill=orange!20, opacity=0.15]  (ap4) circle (1.8cm);

\draw[apedge] (ap1) -- (ap2);
\draw[apedge] (ap1) -- (ap3);
\draw[apedge] (ap1) -- (ap4);
\draw[apedge] (ap2) -- (ap3);
\draw[apedge] (ap2) -- (ap4);
\draw[apedge] (ap3) -- (ap4);

\node[stanode, fill=green] (s1a) at (-2.3, 2.4) {};
\node[stanode, fill=green] (s1b) at (-2.3, 1.0) {};
\draw[staedge]       (ap1) -- (s1a);
\draw[staedgedotted] (ap1) -- (s1b);

\node[stanode, fill=blue] (s2a) at (2.7, 3.0) {};
\draw[staedgedotted] (ap2) -- (s2a);

\node[stanode, fill=red] (s3a) at (4.3, 1.0) {};
\node[stanode, fill=red] (s3b) at (4.1, -0.4) {};
\node[stanode, fill=red] (s3c) at (2.6, -1.0) {};
\draw[staedge]       (ap3) -- (s3a);
\draw[staedgedotted] (ap3) -- (s3b);
\draw[staedgedotted] (ap3) -- (s3c);

\node[stanode, fill=orange] (s4a) at (-0.8, -1.5) {};
\node[stanode, fill=orange] (s4b) at (0.8, -1.6) {};
\draw[staedge]       (ap4) -- (s4a);
\draw[staedgedotted] (ap4) -- (s4b);

\coordinate (attrstart) at ($(ap3)!0.56!(s3a)$);
\draw[-{Stealth[length=3pt]}, black!60, semithick] (attrstart) -- (5.8, 0.8);

\node[attrbox, anchor=south west] at (5.8, -0.1) {%
    \texttt{rssi}\,: $0.23$\\
    \texttt{link\_type}\,: \textsf{AP-STA}\\
    \texttt{active}\,: \textsf{True}\\
    \texttt{selected}\,: \textsf{True}\\
    \texttt{mcs}\,: $7$\\
    \texttt{tx\_power}\,: $3$\\
    \texttt{success}\,: $0.92$
};

\end{tikzpicture}
}%
\caption{Example graph of a Wi-Fi network with four \acp{AP} and eight \acp{STA}. Green dash-dotted edges represent \ac{AP}--\ac{AP} interference links; black edges represent \ac{AP}--\ac{STA} transmission links (solid if selected, dotted if inactive). The legend shows the seven-attribute feature vector carried by each edge.}
\label{fig:graph_example}
\end{figure}
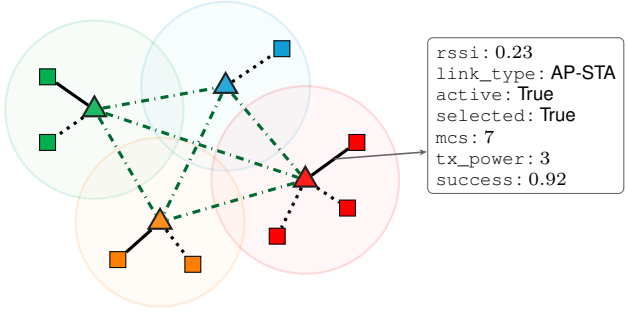

We model the Wi-Fi network as an undirected graph $\mathcal{G} = (\mathcal{V}, \mathcal{E})$, where the vertex set $\mathcal{V} = \mathcal{A} \cup \mathcal{S}$ comprises \acp{AP}~$\mathcal{A}$ and \acp{STA}~$\mathcal{S}$. The edge set~$\mathcal{E}$ contains two types of edges: \ac{AP}--\ac{STA} \emph{transmission links} (one per each associated \ac{AP}--\ac{STA} pair, representing the downlink channel) and \ac{AP}--\ac{AP} \emph{interference edges} (between \acp{AP} whose mutual \ac{RSSI} exceeds the \ac{CCA} threshold of \SI{-82}{dBm}). An example of such a graph is shown in Fig.~\ref{fig:graph_example}. Edges are undirected (i.e., a symmetric channel is assumed) and carry a feature vector defined by a seven-attribute tuple, which we group by their roles:
\begin{itemize}
    \item \textbf{Channel/topology}:
    \begin{itemize}
        \item \texttt{rssi} $\in \mathbb{R}$: received signal strength, standardized as $\tilde{r}_s = (r_s - \mu_\text{RSSI}) / \sigma_\text{RSSI}$ with $\mu_\text{RSSI} = \SI{-51.6}{dBm}$, $\sigma_\text{RSSI} = \SI{14.3}{dB}$ (computed from a training set),
        \item \texttt{link\_type} $\in \{$AP-AP, AP-STA, $\mathsf{N/A}\}$: edge type,
        \item \texttt{active} $\in \{\mathsf{True}, \mathsf{False}, \mathsf{N/A}\}$: indicates whether the \ac{AP}--\ac{STA} pair can be used in \iac{Co-SR} configuration.
    \end{itemize}
    \item \textbf{Scheduling decisions}:
    \begin{itemize}
        \item \texttt{selected} $\in \{\mathsf{True}, \mathsf{False}, \mathsf{N/A}\}$: indicates whether the \ac{AP}--\ac{STA} pair is chosen for transmission,
        \item \texttt{mcs} $\in \{0, 1, \ldots, 13, \mathsf{N/A}\}$: \ac{MCS} index\footnote{Since the final version of the IEEE 802.11bn standard is not released yet, IEEE 802.11be \ac{MCS} indices are assumed.},
        \item \texttt{tx\_power} $\in \{1, 2, 3, 4, \mathsf{N/A}\}$: transmission power level (Table~\ref{tab:sim_params}).
    \end{itemize}
    \item \textbf{Outcome}:
    \begin{itemize}
        \item \texttt{success} $\in [0, 1]$: ratio of \acp{A-MPDU} successfully received at the destination \ac{STA}.
    \end{itemize}
\end{itemize}


\subsection{Autoencoder}
\label{sec:gnn_ae}

\acp{GNN}~\cite{gilmer2017neural} extend deep learning to graph-structured data by iteratively passing learned messages between neighboring elements. We adopt the graph network framework of~\cite{battaglia2018relational}, in which each layer maintains three levels of representation -- edges, nodes, and a global summary vector -- and updates each of them by aggregating information from local neighborhoods through dedicated neural networks. After $L$ such layers, each element's representation captures information from its $L$-hop neighborhood. This makes \acp{GNN} a natural fit for Wi-Fi networks, where interference and scheduling decisions are governed by spatial proximity and local topology. Importantly, the formulation is permutation-equivariant and handles graphs of varying size, allowing a single trained model to generalize across different network deployments.

The \ac{GNN} autoencoder learns a latent representation of Wi-Fi network graphs, jointly encoding both fixed channel/topology attributes and variable scheduling decisions into a latent space so that the \ac{FM} model can generate configurations that are consistent with current channel conditions. The architecture follows \iac{VAE}~\cite{kingma2014auto} adapted to graph-structured data using the \ac{MPNN} framework~\cite{gilmer2017neural}. The latent space is defined \emph{per edge}, i.e., each edge $e$ is mapped to a vector $\mathbf{z}_e \in \mathbb{R}^{d_z}$, preserving the graph topology and allowing the \ac{FM} model to operate on graphs of varying size.\apxonly{ A diagram of the autoencoder is provided in Fig.~\ref{fig:autoencoder_diagram} (Appendix~\ref{sec:architecture_diagrams}).}

\subsubsection{Architecture}
The encoder first projects the flattened edge features into an intermediate embedding space via a linear layer. The embedded graph is then processed by message-passing layers, each implemented as a graph network~\cite{battaglia2018relational} parameterized by residual \acp{MLP} with layer normalization and \ac{GELU} activations~\cite{hendrycks2016bridging}. Messages are aggregated via mean pooling. The encoder projects each edge representation to two vectors $\boldsymbol{\mu}_e$ and $\log \boldsymbol{\sigma}_e^2$ of dimension~$d_z$, defining an approximate posterior $q(\mathbf{z}_e | \mathcal{G}) = \mathcal{N}(\boldsymbol{\mu}_e, \text{diag}(\boldsymbol{\sigma}_e^2))$, where $\boldsymbol{\mu}_e$ is the mean vector and $\text{diag}(\boldsymbol{\sigma}_e^2)$ is the diagonal of the covariance matrix.

The decoder takes a latent graph with edge features $\mathbf{z}_e \sim q(\mathbf{z}_e | \mathcal{G})$ (sampled via the reparameterization trick of~\cite{kingma2014auto}) and processes it through message-passing layers with the same architecture as the encoder. The final edge representations are passed through attribute-specific fully-connected output heads that reconstruct the original attributes.

\subsubsection{Loss Function}
For categorical attributes with $\mathsf{N/A}$ support, the decoder outputs $C$ logits $\mathbf{l} \in \mathbb{R}^C$, where $\mathbf{l}_{1:C-1}$ are the valid-class logits and $l_C$ is the $\mathsf{N/A}$ logit. These parametrize a mixture: a Bernoulli gate decides whether the attribute is $\mathsf{N/A}$, and a categorical over the $C{-}1$ valid classes covers the remaining cases. The \ac{NLL} of this mixture\apxonly{ (derived in Appendix~\ref{sec:na_loss_derivation})} gives the \emph{N/A-gated categorical loss}:
\begin{equation}
    \mathcal{L}_{\text{mix}}(\mathbf{l}, \mathbf{y}) = \mathcal{L}_{\text{BCE}}(l_C, y_C) + (1 - y_C)\,\mathcal{L}_{\text{CE}}(\mathbf{l}_{1:C-1}, \mathbf{y}_{1:C-1}),
    \label{eq:mixture_loss}
\end{equation}
where $\mathcal{L}_{\text{BCE}}$ is the binary cross-entropy and $\mathcal{L}_{\text{CE}}$ is the categorical cross-entropy. When $y_C = 1$, only the gate is trained; when $y_C = 0$, both the gate and the valid-class head contribute. For continuous \texttt{rssi} and \texttt{success} attributes, \ac{MSE} and sigmoid binary cross-entropy losses are used, respectively.

Each attribute-specific loss is thus a proper \ac{NLL} loss under the respective decoder marginal distribution $p_\theta(\mathbf{x}_k | \mathbf{z})$.
The total loss is a weighted negative \ac{ELBO}, where the weights per-attribute account for the different scales and importance of the seven attributes:
\begin{equation}
    \mathcal{L} = \sum_{k=1}^{7} \mathcal{L}_k + \lambda_{\text{KL}} \, D_{\text{KL}}\!\left(q_\phi(\mathbf{z} | \mathcal{G}) \| \mathcal{N}(\mathbf{0}, \mathbf{I})\right),
    \label{eq:gnn_loss}
\end{equation}
where $\mathcal{L}_k$ are per-attribute loss functions and $\lambda_{\text{KL}}$ controls the \ac{KL} regularization strength~\cite{kingma2014auto}.

\subsection{Flow Matching Generative Model}
\label{sec:flow_matching}

The \ac{FM} model learns to generate \ac{Co-SR} configurations using continuous flow matching~\cite{lipman2023flow}, a state-of-the-art generative approach that produces high-quality samples in very few steps, training a velocity field that defines a continuous-time flow in the latent space of the autoencoder.

\subsubsection{Problem Formulation}
Given a context of $n$ recent network observations obtained by executing random \ac{Co-SR} configurations and recording the resulting network state, the goal is to generate a new configuration that is consistent with the observed network conditions. Random (rather than optimized) context configurations are used so that the model does not depend on a particular scheduling policy and can generalize from policy-free probing of the environment.

Let $\mathbf{h} = (\mathbf{h}_1, \ldots, \mathbf{h}_n)$ denote the encoded context, $\mathbf{x} \in \mathbb{R}^{|\mathcal{E}| \times d_z}$ the target latent configuration, and $\mathbf{z} \sim \mathcal{N}(\mathbf{0}, \mathbf{I})$ a noise sample.
We define the linear interpolation path
\begin{equation}
    \mathbf{x}_t = (1-t)\,\mathbf{z} + t\,\mathbf{x}, \quad t \in [0, 1),
    \label{eq:interpolation}
\end{equation}
with the corresponding target velocity field $\mathbf{v}_t = \mathbf{x} - \mathbf{z}$.
The model $v_\theta$ is trained to predict this velocity conditioned on the context by minimizing
\begin{equation}
    \mathcal{L}_{\text{FM}} = \mathbb{E}_{t, \mathbf{z}, \mathbf{x}}\left[\left\| v_\theta(\mathbf{x}_t, t, \mathbf{h}) - \mathbf{v}_t \right\|^2\right],
    \label{eq:fm_loss}
\end{equation}
where $t \sim \mathcal{U}(0, 1)$.

We maintain an \ac{EMA} of the model parameters with a decay rate $\beta_{\text{EMA}} = 0.999$, and use the \ac{EMA} parameters for inference and validation.\apxonly{ The training procedure is illustrated in Fig.~\ref{fig:fm_diagram} (Appendix~\ref{sec:architecture_diagrams}).}

\subsubsection{Architecture}
The $n$ context configurations are encoded by the frozen \ac{GNN} encoder into latent vectors and stacked along a temporal axis. During training, the target configuration is appended as the last slice and the result is flattened into a single vector of dimension $(n+1)d_z$ per edge. The velocity network $v_\theta$ employs \ac{MPNN} with transformer blocks~\cite{vaswani2017attention}. A sinusoidal time embedding~\cite{ho2020denoising} of the timestep~$t$ is calculated and added to the edge features. Within each transformer, attention is masked so that features belonging to different graphs do not attend to each other. A final linear projection maps the edge features back to the latent dimension~$d_z$.

\subsubsection{Sampling via Euler's Method}
The samples are generated by initializing $\mathbf{x}_0 = \mathbf{z} \sim \mathcal{N}(\mathbf{0}, \mathbf{I})$ and solving the \ac{ODE} $\mathrm{d}\mathbf{x}/\mathrm{d}t = v_\theta(\mathbf{x}_t, t, \mathbf{h})$ from $t=0$ (noise) to $t=1$ (data) using the Euler method.
Given a step size $\Delta t = 1/S$, the update at each step is:
\begin{equation}
    \mathbf{x}_{t+\Delta t} = \mathbf{x}_t + v_\theta(\mathbf{x}_t, t, \mathbf{h}) \,\Delta t.
\label{eq:euler}
\end{equation}
The Euler method requires only one model evaluation per step, enabling fast generation. Empirical results are consistent with a smooth velocity field: first-order integration with $S{=}6$ steps achieves high-quality results\apxonly{ (Appendix~\ref{sec:fm_ablations})}.

\subsection{Surrogate Data Rate Predictor}
\label{sec:surrogate}

During inference, a digital twin of the real network is generally unavailable, so candidate configurations cannot be evaluated by simulation. The surrogate model addresses this by providing a data rate predictor, trained offline on simulated data, enabling fast candidate evaluation without access to the real network or its digital twin.\apxonly{ The surrogate architecture is shown in Fig.~\ref{fig:surrogate_diagram} (Appendix~\ref{sec:architecture_diagrams}).}

\subsubsection{Architecture}
The surrogate model takes the embedded graph representation as input and predicts the expected effective data rate. The architecture mirrors the backbone of the \ac{FM} model, and the final global features are projected to the parameters of a Gaussian mixture model following the \ac{MDN}~\cite{bishop1994mdn} framework.

\subsubsection{Mixture Density Network}
The output layer produces $3m$ values per graph, where $m$ is the number of mixture components.
These are split into mixture logits $\boldsymbol{\alpha} \in \mathbb{R}^m$, means $\boldsymbol{\mu} \in \mathbb{R}^m$, and scales $\boldsymbol{\sigma} \in \mathbb{R}_+^m$. The weights of the mixture are obtained as $w_k = \mathrm{softmax}(\boldsymbol{\alpha})_k$. The target is the aggregate data rate, defined as $\sum_{e \in \mathcal{E}_{\text{sel}}} r_e \cdot s_e$ where $\mathcal{E}_{\text{sel}}$ is the set of selected edges, $r_e$ is the PHY-layer data rate of edge~$e$, and $s_e$ is the observed success probability. For the sake of training stability, the data rate is standardized as $\tilde{r}_e = (r_e - \mu_{\text{rate}}) / \sigma_{\text{rate}}$ with $\mu_{\text{rate}} = 657.2$~Mb/s and $\sigma_{\text{rate}} = 462.2$~Mb/s, resulting from a training set. 

The model is trained by minimizing the \ac{NLL} of the standardized data rate under the predicted mixture~\cite{bishop1994mdn}:
\begin{equation}
    \mathcal{L}_{\text{MDN}} = -\log \sum_{k=1}^{m} w_k \cdot f_\mathcal{N}(\tilde{r}_e \mid \mu_k, \sigma_k^2).
    \label{eq:mdn_loss}
\end{equation}

\subsubsection{Candidate Selection}
The mixture density output is used for candidate selection. Given the predicted mixture weights $w_k$ and component parameters, the expected data rate is $\bar{r}_e = \sum_k w_k \mu_k$. Candidates are ranked by their expected data rate, and the top-scoring configurations are retained.

\subsection{Full Inference Pipeline}
\label{sec:inference}

During inference, the pipeline components are combined as illustrated in Fig.~\ref{fig:inference_pipeline}:
\begin{enumerate}
    \item \textbf{Context encoding.} Given the current scenario and $n$ recent random network observations, each observation is converted to a graph and encoded by the frozen \ac{GNN} encoder.
    \item \textbf{Candidate generation.} The encoded context is batched into $N_{\text{cand}}$ copies, each paired with an independent noise sample, and the \ac{FM} model generates $N_{\text{cand}}$ candidate configurations in parallel by solving \ac{ODE} with the Euler method.
    \item \textbf{Evaluation and selection.} All $N_{\text{cand}}$ candidates are scored by the surrogate model, and the top-$k$ configurations with the highest expected data rate are retained.
    \item \textbf{Decoding.} Latent vectors of the top-$k$ candidates are decoded by the decoder and the attributes are extracted to obtain a valid scheduling decision (i.e., exactly one \ac{STA} per active \ac{AP}), with \ac{MCS} index and transmission power level sampled from categorical distributions to maintain diversity across \ac{Co-SR} configuration candidates.
    \item \textbf{Execution.} The selected configurations are applied to the network for the next \acp{TXOP}.
\end{enumerate}

Drawing multiple \ac{Co-SR} configuration candidates in parallel is motivated by the analysis in~\cite{liang2025gdsg}. For a generative model that has learned the underlying solution distribution, the probability of at least one of $N_{\text{cand}}$ independent samples falling within an $\epsilon$-neighborhood of the optimal solution increases as $1 - \bigl[1 - (1-\epsilon)^D\bigr]^{N_{\text{cand}}}$, where $D$ is the dimensionality of the solution space. This probability approaches~1 for sufficiently large $N_{\text{cand}}$, so even when individual samples are suboptimal, parallel sampling combined with surrogate scoring enables the pipeline to approximate the optimum  with high probability.

Because channel conditions and associations change much slower than individual \acp{TXOP}, a well-chosen \ac{Co-SR} configuration can be reused across many consecutive \acp{TXOP} before recalculation is needed. Therefore, the proposed pipeline produces a set of \ac{Co-SR} configurations that is reused in consecutive \acp{TXOP} until conditions change (e.g., new associations or significant performance degradation). The \ac{Co-SR} configuration candidates can be applied in a round-robin fashion or scheduled with an external algorithm, e.g.,~\cite{nunez2025deep}, providing temporal diversity or reducing latency.

\section{Implementation}
\label{sec:implementation}

All experiments use \texttt{mapc\_sim}~\cite{wojnar2025ieee}, an open-source JAX-based IEEE~802.11 Co-SR simulator; simulation parameters are given in Table~\ref{tab:sim_params}\apxonly{ and the full software stack in Appendix~\ref{sec:architecture_diagrams}}.

\begin{table}[t]
    \caption{Simulation parameters used in all experiments.}
    \label{tab:sim_params}
    \centering
    \small
    \begin{tabular}{ll}
        \toprule
        Parameter & Value \\
        \midrule
        Band & \SI{5}{\giga\hertz} \\
        PHY & IEEE~802.11be \\
        Channel width & \SI{80}{\mega\hertz} \\
        Spatial streams & 1, SISO \\
        MCS values & \{0, 1, \ldots, 13\} \\
        Transmission power levels & \{16, 13, 10, 7\}~\si{dBm} \\
        Path loss model & TGax enterprise~\cite{tgax} \\
        Multipath fading & Additive white Gaussian noise \\
        Noise floor & \SI{-94}{dBm} \\
        CCA threshold & \SI{-82}{dBm} \\
        Frame size & \SI{1500}{\byte} \\
        TXOP limit & \SI{5.484}{\milli\second} \\
        Traffic & Full buffer, downlink \\
        \bottomrule
    \end{tabular}
\end{table}

\begin{figure}[t]
\centering
\scalebox{1.0}
{
\begin{tikzpicture}[
    scale=0.5,
    apnode/.style={regular polygon, regular polygon sides=3, draw, inner sep=0pt, minimum size=7pt},
    stanode/.style={rectangle, draw, inner sep=0pt, minimum size=4pt},
    wall/.style={thick, black!80}
]

\begin{scope}[shift={(0,0)}]
    
\node[anchor=south west] at (-1, -1.8) {\small\textbf{(a) Residential topology}};
    
    \draw[wall] (0,0) rectangle (6,6);
    \draw[wall] (3,0) -- (3,6);
    \draw[wall] (0,3) -- (6,3);
    
    
\node[font=\scriptsize, text=black!70] at (1.5, -0.4) {$X$ [m]};
    
\node[font=\scriptsize, text=black!70, rotate=90] at (-0.4, 1.5) {$Y$ [m]};

    
\node[apnode, fill=green] (ap1) at (1.2, 1.8) {};
    
\node[stanode, fill=green] at (0.5, 0.5) {};
    
\node[stanode, fill=green] at (2.2, 0.8) {};
    
\node[stanode, fill=green] at (0.8, 2.5) {};
    
\node[stanode, fill=green] at (2.5, 2.2) {};

    
\node[apnode, fill=blue] (ap2) at (4.5, 1.2) {};
    
\node[stanode, fill=blue] at (3.5, 0.6) {};
    
\node[stanode, fill=blue] at (5.5, 0.9) {};
    
\node[stanode, fill=blue] at (3.8, 2.4) {};
    
\node[stanode, fill=blue] at (5.2, 2.1) {};

    
\node[apnode, fill=red] (ap3) at (1.8, 4.5) {};
    
\node[stanode, fill=red] at (0.6, 3.5) {};
    
\node[stanode, fill=red] at (2.4, 3.8) {};
    
\node[stanode, fill=red] at (0.9, 5.5) {};
    
\node[stanode, fill=red] at (2.1, 5.2) {};

    
\node[apnode, fill=orange] (ap4) at (4.2, 4.8) {};
    
\node[stanode, fill=orange] at (3.4, 3.6) {};
    
\node[stanode, fill=orange] at (5.6, 4.1) {};
    
\node[stanode, fill=orange] at (3.9, 5.4) {};
    
\node[stanode, fill=orange] at (5.1, 5.7) {};
\end{scope}

\begin{scope}[shift={(8,0)}]

\node[anchor=south west] at (0, -1.8) {\small\textbf{(b) Enterprise topology}};

    \draw[dashed, gray] (1.5, 0) -- (1.5, 6);
    \draw[dashed, gray] (4.5, 0) -- (4.5, 6);
    \draw[dashed, gray] (0, 1.5) -- (6, 1.5);
    \draw[dashed, gray] (0, 4.5) -- (6, 4.5);


\node[font=\scriptsize, text=black!70] at (3, -0.4) {$X$ [m]};

\node[font=\scriptsize, text=black!70, rotate=90] at (-0.4, 3) {$Y$ [m]};


\node[apnode, fill=green] (eap1) at (1.5, 1.5) {};

    \node[stanode, fill=green] at (0.3, 1.8) {};
    \node[stanode, fill=green] at (1.9, 0.3) {};
    \node[stanode, fill=green] at (2.6, 2.7) {};


\node[apnode, fill=blue] (eap2) at (4.5, 1.5) {};

    \node[stanode, fill=blue] at (3.4, 2.1) {};
    \node[stanode, fill=blue] at (5.7, 0.4) {};
    \node[stanode, fill=blue] at (4.1, 0.2) {};
    \node[stanode, fill=blue] at (5.3, 2.6) {};


\node[apnode, fill=red] (eap3) at (1.5, 4.5) {};

    \node[stanode, fill=red] at (0.4, 5.1) {};
    \node[stanode, fill=red] at (2.7, 4.0) {};
    \node[stanode, fill=red] at (1.1, 3.4) {};
    \node[stanode, fill=red] at (2.1, 5.7) {};


\node[apnode, fill=orange] (eap4) at (4.5, 4.5) {};

    \node[stanode, fill=orange] at (3.3, 5.4) {};
    \node[stanode, fill=orange] at (5.6, 3.5) {};
    \node[stanode, fill=orange] at (4.0, 3.8) {};
    \node[stanode, fill=orange] at (5.8, 5.2) {};
    \node[stanode, fill=orange] at (3.7, 4.8) {};

\end{scope}

\end{tikzpicture}
}
\caption{Example topologies of the scenarios: (a) residential (structured room layout with walls), (b) enterprise (regular AP grid without walls; dashed lines indicate room boundaries).} 
\label{fig:exemplary_topologies}
\end{figure}
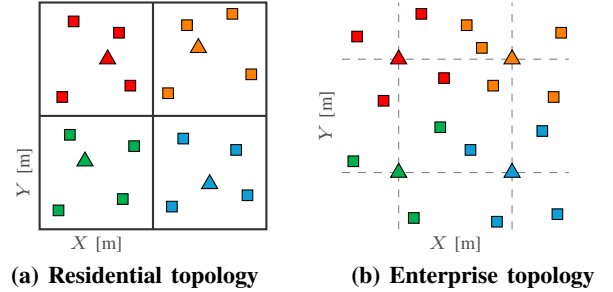

\subsection{Training Scenarios}
\label{sec:dataset}

Network topologies are drawn from a set of parameterized scenarios that can be instantiated with varying \ac{AP}/\ac{STA} counts, spatial layouts, and radio parameters. Moreover, different baseline models are used to generate \ac{Co-SR} configurations for training and validation datasets.

To train models that generalize across diverse network deployments, we use many parameterized scenario types that span a wide range of topologies. These include (i) small topologies (2~\acp{AP} with variable inter-\ac{AP} distances of 1--50\,m), (ii) structured \textit{residential scenarios} from~\cite{tgax} (rooms arranged in grids from 2x2 to 5x2 square rooms with 1 \ac{AP} and 1--6 \acp{STA} per room, 5--20\,m room width, with walls providing additional attenuation), (iii) open-space random deployments (2--10~\acp{AP}, 1--5~\acp{STA} per \ac{AP}, in square areas with a width of 20--100\,m), (iv) dense co-located deployments (2--10~\acp{AP}, 1--5~associations each), (v) \textit{indoor small \acp{BSS} scenarios} from~\cite{tgax}, and (vi) challenging interference patterns (hidden node and flow-in-the-middle topologies). Network sizes range from 2~\acp{AP} with 2~\acp{STA} to 10~\acp{AP} with up to 50~\acp{STA}.

During dataset generation, scenarios are sampled with a probability proportional to the number of their configurable parameters, ensuring that scenarios with larger parameter spaces (which exhibit greater variability and thus require more samples for adequate coverage) contribute proportionally more training data. For each sampled scenario type, a realization is drawn by uniformly sampling all parameters within their respective ranges.

\subsection{Baseline Configuration Algorithms}
\label{sec:baselines_algo}
For each scenario realization, \ac{Co-SR} configurations are generated using four complementary algorithms:
\begin{itemize}
    \item \textbf{Random:} uniform random sampling of \ac{AP}--\ac{STA} pairs, transmission power levels, and \ac{MCS} index values.
    \item \textbf{H-MAB:} \acp{H-MAB} proposed in~\cite{wojnar2025csr}. Three levels of agents sequentially select active \acp{AP}, recipient \acp{STA}, and transmission power, while \ac{MCS} values are selected by an oracle.
    \item \textbf{F-Optimal:} max-min fairness upper bound solver, following the formulation in~\cite{wojnar2025csr}.
    \item \textbf{T-Optimal:} maximum aggregate data rate upper bound solver, optimized for total throughput~\cite{wojnar2025csr}.
\end{itemize}
This mixture ensures that the training data covers both optimal configurations (F-Optimal, T-Optimal) and the broader configuration landscape (Random, \ac{H-MAB}), allowing the generative model to learn the underlying structure of viable \ac{Co-SR} configurations. No \ac{DRL} baseline is included: the only directly relevant \ac{DRL} scheme for \ac{Co-SR} configuration generation, SAIOC~\cite{jung2026coordinated}, does not provide a public implementation, and we found no other reproducible \ac{DRL} baseline addressing the same problem.
Additionally, we do not compare directly against legacy IEEE 802.11 operation because it is consistently outperformed by H-MAB in Co-SR scenarios~\cite{wojnar2025csr}.
Finally, we note that only FM4WiFi provides MCS selection, while the other baselines assume perfect MCS values. Therefore, to isolate scheduling quality, we also report FM4WiFi with the same oracle \ac{MCS} to enable a more direct and fair comparison with the state-of-the-art baselines.

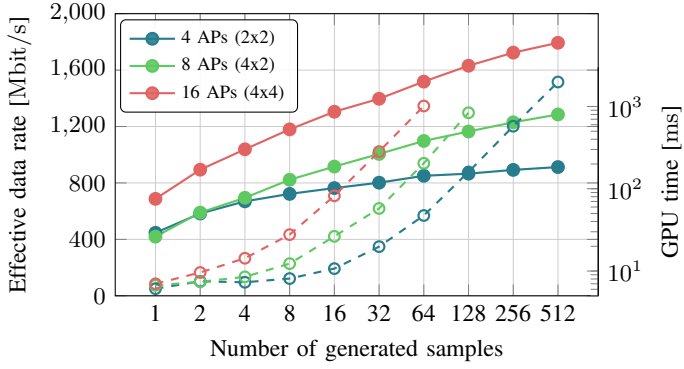
\begin{figure}[t]
\centering
\scalebox{1.0}
{
\begin{tikzpicture}
\colorlet{cap4}{palB}
\colorlet{cap8}{palD}
\colorlet{cap16}{palF}

\begin{axis}[
  name=mainax,
  width=0.9\columnwidth,
  height=0.6\columnwidth,
  xlabel={Number of generated samples},
  ylabel={Effective data rate [\si{\mega\bit\per\second}]},
  ymin=0, ymax=2000,
  ytick distance=400,
  xmode=log,
  log basis x=2,
  xtick={1,2,4,8,16,32,64,128,256,512},
  xticklabels={1,2,4,8,16,32,64,128,256,512},
  grid=both,
  major grid style={line width=0.2pt, draw=gray!40},
  minor grid style={line width=0.1pt, draw=gray!20},
  tick label style={font=\small},
  label style={font=\small},
  every axis plot/.append style={thick, mark size=2.2pt},
  legend style={at={(0.01,0.98)}, anchor=north west, draw=black, fill=white, fill opacity=0.9, text opacity=1, font=\scriptsize, rounded corners=2pt},
  legend columns=1,
  legend cell align=left,
]
\addplot[color=cap4, solid, mark=*] coordinates {(1,446.67) (2,583.43) (4,669.50) (8,722.33) (16,763.61) (32,801.82) (64,850.62) (128,865.80) (256,892.53) (512,912.62)};
\addlegendentry{4 APs (2x2)}
\addplot[color=cap8, solid, mark=*] coordinates {(1,419.49) (2,589.87) (4,694.75) (8,822.69) (16,916.03) (32,1005.53) (64,1097.68) (128,1164.81) (256,1229.76) (512,1285.46)};
\addlegendentry{8 APs (4x2)}
\addplot[color=cap16, solid, mark=*] coordinates {(1,687.14) (2,894.07) (4,1038.25) (8,1179.86) (16,1304.75) (32,1396.78) (64,1518.29) (128,1630.89) (256,1723.78) (512,1793.38)};
\addlegendentry{16 APs (4x4)}
\end{axis}

\begin{axis}[
  width=0.9\columnwidth,
  height=0.52\columnwidth,
  xmode=log,
  log basis x=2,
  xtick={1,2,4,8,16,32,64,128,256,512},
  axis x line=none,
  axis y line*=right,
  ylabel={GPU time [\si{\milli\second}]},
  ymin=5, ymax=3000,
  ymode=log,
  tick label style={font=\small},
  label style={font=\small},
]
\addplot[color=cap4, dashed, mark=o, mark options={solid}] coordinates {(1,6.1) (2,7.5) (4,7.3) (8,8.1) (16,10.7) (32,19.8) (64,47.3) (128,160.0) (256,575.7) (512,1983.7)};
\addplot[color=cap8, dashed, mark=o, mark options={solid}] coordinates {(1,6.7) (2,7.4) (4,8.5) (8,12.3) (16,26.4) (32,57.6) (64,203.9) (128,838.9)};
\addplot[color=cap16, dashed, mark=o, mark options={solid}] coordinates {(1,7.0) (2,9.6) (4,14.3) (8,27.7) (16,82.0) (32,283.2) (64,1013.1)};
\end{axis}
\end{tikzpicture}
}
\caption{FM4WiFi sample count scaling for residential scenarios: solid curves (left axis) show effective data rate; dashed curves (right axis) -- inference time.}
\label{fig:samples_vs_throughput_time}
\end{figure}

\subsection{Dataset Composition}
For each baseline algorithm, we generate a training set (1,000 scenario realizations) and a validation set (200 realizations). Each realization contributes 5--50 \ac{Co-SR} configurations depending on the algorithm (5 for T-Optimal, 30 for Random and F-Optimal, 50 for \ac{H-MAB}), yielding $\sim 138{,}000$ examples in total. The random dataset is excluded from \ac{FM} training because it contains degraded configurations that would bias the generative model. The autoencoder is trained on all four baseline datasets, including the random \ac{Co-SR} configurations. The surrogate model is trained separately on configurations generated by the trained \ac{FM} model and evaluated in the simulator to obtain ground-truth data rates.

\section{Evaluation}

We first characterize the evaluated scenario families and inference-time design choices, then benchmark FM4WiFi against established static-scenario baselines, and analyze periodic regeneration under mobility.

Fig.~\ref{fig:exemplary_topologies} illustrates the scenario families used throughout this section. Residential topologies \cite{tgax} make it straightforward to build realistic large-scale indoor deployments with structured propagation constraints (i.e., walls). In contrast, enterprise topologies arrange \acp{AP} on a regular grid with stations randomly placed within each cell, but without walls. We additionally evaluate the proposed approach on open-space topologies, in which \acp{AP} are placed randomly within a \SI{75}{\meter} $\times$ \SI{75}{\meter} area without walls, and station positions are drawn from a normal distribution centered at their associated \ac{AP} with variable standard deviation. Each open-space topology has 2--5~\acp{AP} and 3--5 stations per \ac{AP}. Halfway through each open-space simulation, all station positions are redrawn to introduce a sudden topology change.

\subsection{Scalability and Number of Samples}
\label{sec:scalability}

\begin{figure}[t]
\centering
\scalebox{1.0}
{
\begin{tikzpicture}
\colorlet{cap4}{palB}
\colorlet{cap8}{palD}
\colorlet{cap16}{palF}

\begin{axis}[
  width=0.9\columnwidth,
  height=0.6\columnwidth,
  xlabel={Number of top selected configs},
  ylabel={Effective data rate [\si{\mega\bit\per\second}]},
  ymin=0, ymax=2000,
  ytick distance=400,
  xmode=log,
  log basis x=2,
  xtick={1,2,4,8,16,32,64,128},
  xticklabels={1,2,4,8,16,32,64,128},
  grid=both,
  major grid style={line width=0.2pt, draw=gray!40},
  minor grid style={line width=0.1pt, draw=gray!20},
  tick label style={font=\small},
  label style={font=\small},
  every axis plot/.append style={thick, mark size=2.2pt},
  legend style={at={(0.01,0.98)}, anchor=north west, draw=black, fill=white, fill opacity=0.9, text opacity=1, font=\scriptsize, rounded corners=2pt, column sep=2pt, /tikz/every even column/.append style={column sep=4pt}},
  legend columns=3,
  legend cell align=left,
]
\addplot[color=cap4, solid, mark=*] coordinates {(1,865.80) (2,849.26) (4,818.39) (8,778.15) (16,745.04) (32,719.57) (64,645.99) (128,445.78)};
\addlegendentry{2x2}
\addplot[color=cap8, solid, mark=*] coordinates {(1,1164.81) (2,1124.30) (4,1056.51) (8,976.42) (16,880.53) (32,779.77) (64,648.41) (128,428.71)};
\addlegendentry{4x2}
\addplot[color=cap16, solid, mark=*] coordinates {(1,1630.89) (2,1555.52) (4,1478.19) (8,1371.60) (16,1254.99) (32,1112.47) (64,939.37) (128,674.30)};
\addlegendentry{4x4}
\end{axis}

\begin{axis}[
  width=0.9\columnwidth,
  height=0.52\columnwidth,
  xmode=log,
  log basis x=2,
  xtick={1,2,4,8,16,32,64,128},
  axis x line=none,
  axis y line*=right,
  ylabel={Jain's fairness index},
  ymin=0.00,
  ymax=1.00,
  tick label style={font=\small},
  label style={font=\small},
]
\addplot[color=cap4, dashed, mark=o, mark options={solid}] coordinates {(1,0.1512) (2,0.2147) (4,0.2711) (8,0.3265) (16,0.4319) (32,0.5782) (64,0.6800) (128,0.7291)};
\addplot[color=cap8, dashed, mark=o, mark options={solid}] coordinates {(1,0.0979) (2,0.1364) (4,0.1962) (8,0.2765) (16,0.3540) (32,0.4491) (64,0.5372) (128,0.6116)};
\addplot[color=cap16, dashed, mark=o, mark options={solid}] coordinates {(1,0.0733) (2,0.1122) (4,0.1661) (8,0.2286) (16,0.2926) (32,0.3561) (64,0.4172) (128,0.4931)};
\end{axis}
\end{tikzpicture}
}
\caption{FM4WiFi top-$k$ selection scaling for residential scenarios with 128 generated candidates: solid curves (left axis) show effective data rate, and dashed curves (right axis) show Jain's fairness index.}
\label{fig:top_configs_vs_throughput_fairness}
\end{figure}
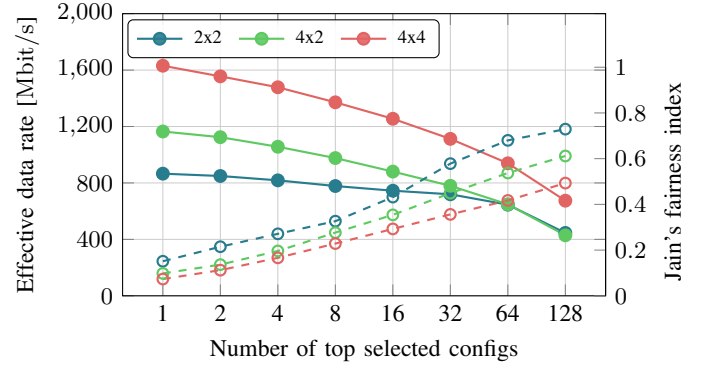

Fig.~\ref{fig:samples_vs_throughput_time} evaluates the effect of candidate-set size under a fixed top-1 selection policy. To isolate the effectiveness of FM sampling, generated candidates are evaluated directly in the simulator rather than through the surrogate model. Each data point is averaged over three realizations of each scenario type with different random node placements. Effective data rate grows roughly linearly with the number of samples, reaching diminishing returns beyond 128 candidates for small networks. The reported GPU time covers only the compiled inference pipeline; input graph construction on the CPU is excluded as it is negligible relative to GPU overhead. GPU time scales quadratically with the number of samples and the network size, staying below \SI{50}{\milli\second} for 64 candidates in the 2x2 scenario and approximately \SI{1}{\second} for the 4x4 scenario. The quadratic scaling stems from the transformer's self-attention mechanism. Notably, generating all candidates in a single large batch is suboptimal: since graphs within a batch are masked from one another to keep their computations independent, the joint attention matrix becomes highly sparse, wasting compute. Generating candidates in smaller batches is therefore more efficient in practice. We adopt 128 samples as the default operating point, balancing throughput gains against GPU latency.

Next, we analyze the effect of selecting the top-$k$ candidates instead of the single highest-ranked one (Fig.~\ref{fig:top_configs_vs_throughput_fairness}). Since not all stations are involved in a single \ac{Co-SR} candidate, top-1 selection can over-focus on a subset of stations. Keeping the top-$k$ provides a practical fairness--data rate trade-off: data rate decreases moderately while Jain's fairness index improves substantially (e.g., from 0.15 to 0.73 for the 2x2 scenario). We set $k=8$ as the default since it typically covers all stations while preserving a strong data rate.

\begin{figure}[t]
\centering
\begin{tikzpicture}
\colorlet{ctopt}{palA}
\colorlet{cfopt}{palB2}
\colorlet{chmab}{palD}
\colorlet{cfm}{palF}
\colorlet{cfmo}{palH}
\definecolor{chmab10}{RGB}{226,245,226}
\definecolor{chmab20}{RGB}{211,240,212}
\definecolor{chmab30}{RGB}{196,235,198}
\definecolor{chmab40}{RGB}{182,230,183}
\definecolor{chmab50}{RGB}{167,226,169}
\definecolor{chmab60}{RGB}{153,221,154}
\definecolor{chmab70}{RGB}{138,216,140}
\definecolor{chmab80}{RGB}{123,211,126}
\definecolor{chmab90}{RGB}{109,206,111}
\colorlet{chmab100}{palD}
 \tikzset{
        every pin/.style={rectangle,rounded corners=5pt,font=\scriptsize},
        small dot/.style={fill=black,circle,scale=0.3},
    }
\begin{axis}[
  width=\columnwidth,
  height=0.8\columnwidth,
  xlabel={Number of APs in scenario},
  ylabel={Effective data rate [\si{\mega\bit\per\second}]},
  ymin=0, ymax=4750,
  ytick distance=1000,
  xtick={3.5,6,9,12,16,20,25,30},
  xticklabels={2x2,2x3,3x3,3x4,4x4,4x5,5x5,5x6},
  grid=both,
  major grid style={line width=0.2pt, draw=gray!40},
  minor grid style={line width=0.1pt, draw=gray!20},
  enlarge x limits=0.1,
  clip=false,
  tick label style={font=\small},
  xticklabel style={font=\small, align=center},
  label style={font=\small},
  every axis plot/.append style={thick, mark size=2.2pt},
  legend style={
    at={(0.5,0.99)},
    anchor=north,
    draw=black,
    fill=white,
    fill opacity=0.9,
    text opacity=1,
    font=\scriptsize,
    rounded corners=2pt
  },
  legend cell align=left,
  legend columns=3
]
\addplot[color=chmab10, dashed, mark=none, line width=0.9pt, forget plot] coordinates {(3.5,554.865062) (6,419.380015) (9,744.769147) (12,1028.639168) (16,1129.649891)};
\addplot[color=chmab20, dashed, mark=none, line width=0.9pt, forget plot] coordinates {(3.5,690.459519) (6,441.765135) (9,784.908643) (12,1091.056729) (16,1130.350109)};
\addplot[color=chmab30, dashed, mark=none, line width=0.9pt, forget plot] coordinates {(3.5,755.803550) (6,527.330416) (9,815.078775) (12,1132.662327) (16,1142.443472)};
\addplot[color=chmab40, dashed, mark=none, line width=0.9pt, forget plot] coordinates {(3.5,783.132750) (6,625.787746) (9,838.217724) (12,1162.908972) (16,1151.807987)};
\addplot[color=chmab50, dashed, mark=none, line width=0.9pt, forget plot] coordinates {(3.5,795.614880) (6,694.720642) (9,915.122538) (12,1184.947987) (16,1156.949672)};
\addplot[color=chmab60, dashed, mark=none, line width=0.9pt, forget plot] coordinates {(3.5,805.557987) (6,737.369317) (9,1045.278082) (12,1202.031090) (16,1159.737418)};
\addplot[color=chmab70, dashed, mark=none, line width=0.9pt, forget plot] coordinates {(3.5,807.823278) (6,770.768990) (9,1142.098468) (12,1215.395217) (16,1157.541419)};
\addplot[color=chmab80, dashed, mark=none, line width=0.9pt, forget plot] coordinates {(3.5,813.048869) (6,795.430343) (9,1214.227571) (12,1227.572210) (16,1161.039387)};
\addplot[color=chmab90, dashed, mark=none, line width=0.9pt, forget plot] coordinates {(3.5,816.964098) (6,814.549801) (9,1272.197180) (12,1237.252067) (16,1164.945295)};

\addplot[color=ctopt, solid, mark=*, mark options={solid}] coordinates {(3.5,1128.9) (6,1165.0) (9,1633.4) (12,2834.4) (16,3206.7) (20,3567.1)};
\addlegendentry{T-Optimal}

\addplot[color=cfopt, solid, mark=*, mark options={solid}] coordinates {(3.5,801.5038355813325) (6,875.2661630544156) (9,1158.0232544994062) (12,1395.8932573287711) (16,1655.4317356008007) (20,1958.243962961537)};
\addlegendentry{F-Optimal}

\addplot[color=chmab100, solid, mark=*, mark options={solid}] coordinates {(3.5,819.045952) (6,831.202042) (9,1317.857112) (12,1245.538993) (16,1162.109409)};
\addlegendentry{H-MAB}

\addplot[color=cfmo, solid, mark=*, mark options={solid}] coordinates {(3.5,736.0674234135665) (6,699.8786241794309) (9,1057.195363785558) (12,1534.3886761487965) (16,1581.8278172866521) (20,1829.1096827133479) (25,2233.6313594091903) (30,2526.9471416849015)};
\addlegendentry{FM4WiFi (oracle MCS)}

\addplot[color=cfm, solid, mark=*, mark options={solid}] coordinates {(3.5,705.6191876367615) (6,665.3207056892779) (9,662.6880470459519) (12,960.855443107221) (16,943.3670678336981) (20,964.8163977024071) (25,1356.632932166302) (30,1382.0603118161926)};
\addlegendentry{FM4WiFi}

\addlegendimage{empty legend}
\addlegendentry{}

\path[draw=none, fill=chmab10]  (axis cs:14.00,385.0) rectangle (axis cs:15.50,435.0);
\path[draw=none, fill=chmab20]  (axis cs:15.50,385.0) rectangle (axis cs:17.00,435.0);
\path[draw=none, fill=chmab30]  (axis cs:17.00,385.0) rectangle (axis cs:18.50,435.0);
\path[draw=none, fill=chmab40]  (axis cs:18.50,385.0) rectangle (axis cs:20.00,435.0);
\path[draw=none, fill=chmab50]  (axis cs:20.00,385.0) rectangle (axis cs:21.50,435.0);
\path[draw=none, fill=chmab60]  (axis cs:21.50,385.0) rectangle (axis cs:23.00,435.0);
\path[draw=none, fill=chmab70]  (axis cs:23.00,385.0) rectangle (axis cs:24.50,435.0);
\path[draw=none, fill=chmab80]  (axis cs:24.50,385.0) rectangle (axis cs:26.00,435.0);
\path[draw=none, fill=chmab90]  (axis cs:26.00,385.0) rectangle (axis cs:27.50,435.0);
\path[draw=none, fill=chmab100] (axis cs:27.50,385.0) rectangle (axis cs:29.00,435.0);
\draw[black, line width=0.25pt] (axis cs:14.0,385.0) rectangle (axis cs:29.0,435.0);
\node[font=\scriptsize, anchor=south] at (axis cs:22.2,440) {H-MAB stage [\%]};
\draw[black, line width=0.25pt] (axis cs:14,385) -- (axis cs:14,360);
\draw[black, line width=0.25pt] (axis cs:21.5,385) -- (axis cs:21.5,360);
\draw[black, line width=0.25pt] (axis cs:29,385) -- (axis cs:29,360);
\node[font=\footnotesize, anchor=north] at (axis cs:14,373) {0};
\node[font=\footnotesize, anchor=north] at (axis cs:21.3,373) {50};
\node[font=\footnotesize, anchor=north] at (axis cs:29,373) {100};

\addplot[dashed, thick, color=palC] coordinates {(18,0) (18,4750)};
\node [small dot,pin=-180:{Resource limit for H-MAB}]    at (axis cs:18,3450)   {};

\addplot[dashed, thick, color=palB] coordinates {(22,0) (22,4750)};
\node [small dot,pin=-180:{Resource limit for analytical models}]  at (axis cs:22,3700)   {};

\end{axis}
\end{tikzpicture}
\caption{Effective data rate across scenario sizes. Curves are shown for visual guidance. Dashed \ac{H-MAB} curves show intermediate convergence stages; vertical dashed lines -- tractability limits for \ac{H-MAB} and analytical baselines, respectively.}
\label{fig:static_residential}
\end{figure}
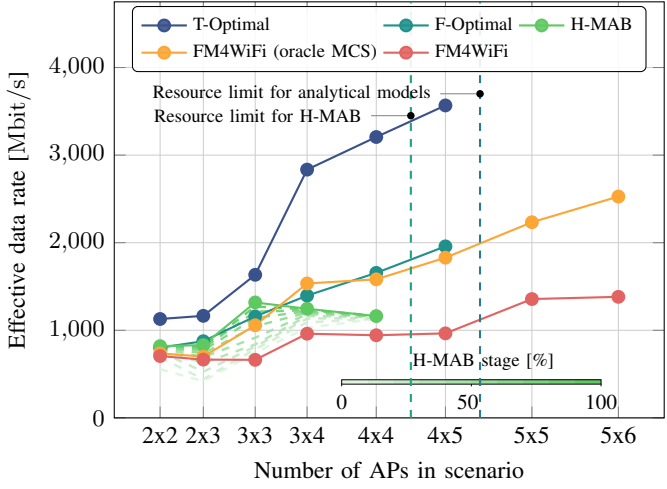

\subsection{Static Scenarios}

All static experiments use residential scenarios with room widths of \SI{10}{\meter}, 1~\ac{AP} and 4~\acp{STA} per room with random placement and topologies ranging from 2x2 to 5x6 apartments. For FM4WiFi, the full inference pipeline (128 candidates, top 8 selected) is executed 32 times per scenario; for each run, the reported value is the mean over 5 simulator evaluations of the selected configurations, and the final result is the mean across all runs. T-Optimal and F-Optimal are deterministic and computed once per scenario. For \ac{H-MAB}, the reported value is the mean over the last 20\% of the run, after convergence or upon reaching the computation budget; analytical baselines (T-Optimal, F-Optimal) are granted up to 12~hours per scenario, while \ac{H-MAB} is additionally limited by memory: at larger scales, agent initialization exceeds available RAM and the method cannot run at all.

Fig.~\ref{fig:static_residential} compares FM4WiFi (with and without oracle MCS) with \ac{H-MAB}, F-Optimal, and T-Optimal. T-Optimal achieves the highest data rate but imposes no fairness constraints, starving all but a few selected stations. F-Optimal enforces equal per-station rates and serves as a more practical upper bound. 
At small scales (2x2--3x3), \ac{H-MAB} outperforms FM4WiFi even with oracle \ac{MCS}, benefiting from online adaptation to a single topology. As the network grows, \ac{H-MAB}'s data rate degrades and it fails to converge within the time budget, while FM4WiFi with oracle \ac{MCS} overtakes it at 12~\acp{AP} (3x4), staying within 7\% of F-Optimal at larger scales. Beyond 16~\acp{AP} (4x4), \ac{H-MAB} becomes intractable, and beyond 20~\acp{AP} (4x5) so do the analytical baselines, while FM4WiFi scales to 30~\acp{AP} (5x6) and beyond.
Wall-clock cost follows the same ordering: in the 4x4 scenario, FM4WiFi generates and scores 128 candidates in \SI{0.62}{\second}, whereas T-Optimal requires \SI{29.7}{\second}, F-Optimal nearly 50~minutes, and \ac{H-MAB} does not converge at all\apxonly{ (Table~\ref{tab:convergence_times})}.

\apxonly{
    Fig.~\ref{fig:random_thr_boxplot} further evaluates small open-space networks (2--5~\acp{AP}), using 24 scenario realizations evaluated twice (i.e., before and after the mid-simulation position is redrawn), producing 48 data points in total. When comparing under equal conditions (oracle \ac{MCS}), FM4WiFi matches \ac{H-MAB} -- the medians are \SI{542}{\mega\bit\per\second} and \SI{505}{\mega\bit\per\second}, respectively, with near-identical distributions. Both methods exhibit high variability across scenario instances, reflecting the diversity of random topologies rather than instability of the algorithms.
}

\apxonly{
    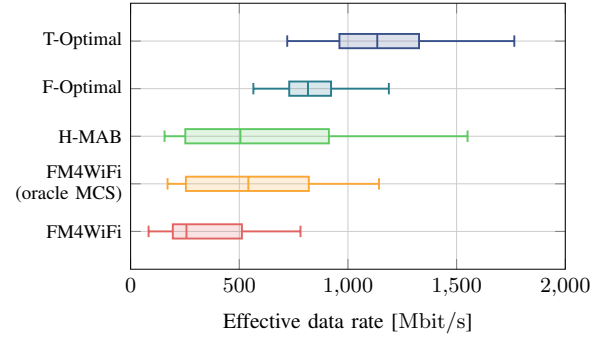
\begin{figure}[t]
    \centering
    \begin{tikzpicture}
\begin{scope}[scale=0.9]
\colorlet{ctoptimal}{palA}
\colorlet{chmab}{palD}
\colorlet{cfoptimal}{palB}
\colorlet{cfm}{palF}
\colorlet{cfmo}{palH}

\begin{axis}[
  width=0.9\columnwidth,
  height=0.62\columnwidth,
  xlabel={Effective data rate [\si{\mega\bit\per\second}]},
  xmin=0.0,
  xmax=2000,
  grid=both,
  major grid style={line width=0.2pt, draw=gray!40},
  minor grid style={line width=0.1pt, draw=gray!20},
  tick label style={font=\small},
  label style={font=\small},
  ytick={1,2,3,4,5},
  yticklabels={FM4WiFi, {FM4WiFi\\(oracle MCS)}, H-MAB, F-Optimal, T-Optimal},
  y tick label style={align=right, font=\footnotesize},
  enlarge y limits=0.15,
  boxplot/draw direction=x,
  boxplot/variable width,
  boxplot/box extend=0.3,
  boxplot/every box/.style={solid},
  boxplot/every whisker/.style={solid},
  boxplot/every median/.style={solid},
  boxplot/every outlier/.style={solid, mark=*, mark size=1.8pt},
]

\addplot+[
  boxplot prepared={draw position=5, median=1134.9498901927, lower quartile=960.8, upper quartile=1327.1, lower whisker=720.6, upper whisker=1765.5},
  draw=ctoptimal, fill=ctoptimal, fill opacity=0.18, thick, mark=*,
] coordinates {};

\addplot+[
  boxplot prepared={draw position=4, median=816.0664147395, lower quartile=729.7971631502, upper quartile=922.303683479, lower whisker=564.9187794792, upper whisker=1188.1785711517},
  draw=cfoptimal, fill=cfoptimal, fill opacity=0.18, thick, mark=*,
] coordinates {};

\addplot+[
  boxplot prepared={draw position=3, median=505.2352297593, lower quartile=251.6849015317, upper quartile=912.2985047411, lower whisker=156.1378555799, upper whisker=1550.3282275711},
  draw=chmab, fill=chmab, fill opacity=0.18, thick, mark=*,
] coordinates {};

\addplot+[
  boxplot prepared={draw position=2, median=541.6301969365, lower quartile=254.3272189551, upper quartile=819.938970186, lower whisker=169.9449535011, upper whisker=1142.9499452954},
  draw=cfmo, fill=cfmo, fill opacity=0.18, thick, mark=*,
] coordinates {};

\addplot+[
  boxplot prepared={draw position=1, median=257.0449261488, lower quartile=194.1419071389, upper quartile=512.2217758479, lower whisker=82.5851340263, upper whisker=781.7047319475},
  draw=cfm, fill=cfm, fill opacity=0.18, thick, mark=*,
] coordinates {};

\end{axis}
\end{scope}
\end{tikzpicture}
    \caption{Effective data rate distribution across open-space scenarios with 2--5 APs.}
    \label{fig:random_thr_boxplot}
    \end{figure}
}

\subsection{Dynamic Scenarios}

\begin{figure}[t]
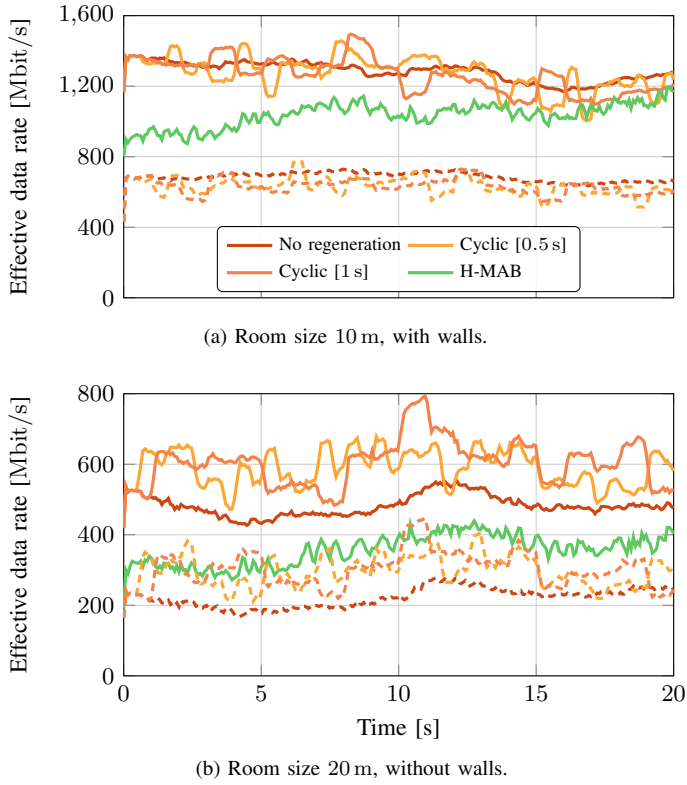

\centering
\subfloat[Room size \SI{10}{\meter}, with walls.]{\input{figures/dynamic_scenarios_res.tikz}\label{fig:dynamic_scenarios_res}}\\
\subfloat[Room size \SI{20}{\meter}, without walls.]{\input{figures/dynamic_scenarios_res_no_walls.tikz}\label{fig:dynamic_scenarios_res_no_walls}}
\caption{Effective data rate over time for the 4x4 residential scenario with mobility. Dashed lines show FM4WiFi with its own \ac{MCS} selection; solid lines -- FM4WiFi with oracle \ac{MCS}.}
\label{fig:dynamic_scenarios_res_merged}
\end{figure}

FM4WiFi is designed for static snapshots: given the current network state, it outputs a \ac{Co-SR} configuration for that moment. Therefore, for dynamic operation, we recommend periodic recomputation. All dynamic results are averaged over five independent runs; instantaneous throughput is reported as a \SI{0.2}{\second} moving average. Stations follow a random waypoint mobility model (with a speed sampled from $U[0.3, 1.5]\,\si{\meter\per\second}$ and a pause -- from $U[0, 5]\,\si{\second}$) in a 4x4 residential layout. We compare a one-shot configuration with cyclic regeneration of the Co-SR configuration at intervals of 0.5, 1, and \SI{2}{\second}. Fig.~\ref{fig:dynamic_scenarios_res} shows results for a room width of \SI{10}{\meter} (with walls), while Fig.~\ref{fig:dynamic_scenarios_res_no_walls} presents results with walls removed and a room width of \SI{20}{\meter}.

In Fig.~\ref{fig:dynamic_scenarios_res}, stations move within their rooms to avoid reassociation. This is the simplest of the dynamic scenarios: walls partition the interference structure, so the link conditions remain largely stable despite mobility.  FM4WiFi (oracle MCS) achieves a data rate around \SIrange{1100}{1400}{\mega\bit\per\second}, confirming that a single generated configuration is sufficient under these conditions. \ac{H-MAB} starts well below FM4WiFi (oracle MCS) and is still visibly converging at $t = \SI{20}{\second}$.
FM4WiFi, despite the decreased performance due to MCS selection, maintains consistency over time.

In Fig.~\ref{fig:dynamic_scenarios_res_no_walls}, stations also stay within the initial room boundaries. However, with walls removed, the propagation environment changes more as stations move, leading to often radical changes in interference patterns. Consequently, periodic regeneration yields significant gains: for FM4WiFi (oracle MCS) cyclic~[\SI{0.5}{\second}] outperforms the no-regeneration baseline by roughly 25\%, at the cost of higher throughput variability. \ac{H-MAB} converges even more slowly in this larger topology, staying well below all FM4WiFi (oracle MCS) curves while achieving only slightly better performance than FM4WiFi with its own MCS selection.\apxonly{ Individual per-seed traces for this scenario are provided in Appendix~\ref{sec:additional_results}.}

\begin{figure}[t]
\centering
\input{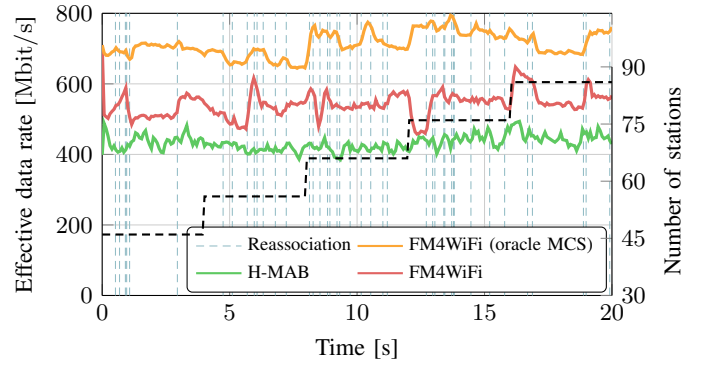}
\caption{Effective data rate for the 3x3 enterprise scenario. Stations move freely and may reassociate; 10 new stations join every \SI{4}{\second} (right axis). Vertical lines mark reassociations.}
\label{fig:dynamic_scenarios_enterprise}
\end{figure}

Finally, Fig.~\ref{fig:dynamic_scenarios_enterprise} evaluates the most demanding enterprise scenario representing managed venues where \acp{AP} are centrally coordinated and the number of connected stations grows over time. This highly demanding and strongly dynamic setup reflects realistic enterprise deployments. The scenario starts with 46~stations; every \SI{4}{\second}, 10~new stations join, reaching 86~by the end. Stations also move freely throughout the area, triggering reassociations when a closer \ac{AP} is found; each such event is marked by a vertical line. FM4WiFi responds to each join or reassociation by immediately running inference on the updated network state. Despite continuous disruptions, FM4WiFi with oracle \ac{MCS} maintains a stable throughput around \SIrange{680}{760}{\mega\bit\per\second}, recovering quickly after each event. \ac{H-MAB}, by contrast, maintains a static action space tied to the current set of associated stations: any join or reassociation requires complete reset and relearning from scratch. Faced with continuous disruptions, \ac{H-MAB} never converges and stays flat around \SIrange{400}{480}{\mega\bit\per\second} throughout the run, well below both FM4WiFi variants.
This challenging scenario demonstrates that FM4WiFi, even without an MCS oracle, can outperform a state-of-the-art solution (H-MAB) operating with perfect MCS selection.

\subsection{Experimental Validation}

\begin{figure}
    \centering
    \includegraphics[width=\linewidth]{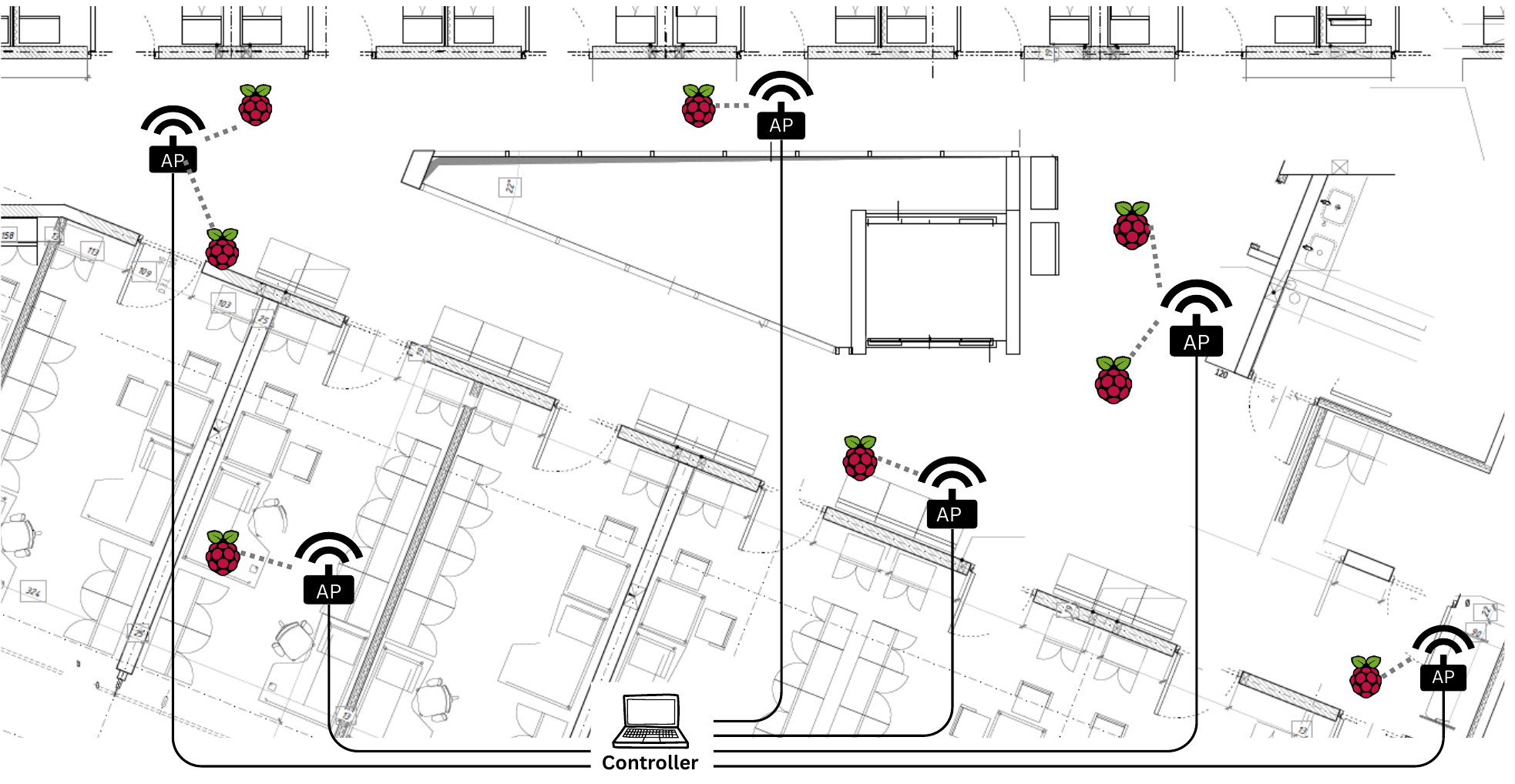}
    \caption{Experimental topology (indoor office).}
    \label{fig:topology}
\end{figure}

For experimental validation, we use an indoor testbed with six APs, eight stations, and an FM4WiFi controller (Fig.~\ref{fig:topology}). We use FM4WiFi sampling 128 candidates and top-4 selection. The model is not retrained, so it operates zero-shot on a previously unseen topology and, for the first time, in a real-world environment.
The APs are laptops with Atheros AR9271 cards running Modwifi~\cite{modwifi} for low-level control over frame transmissions; the stations are Raspberry Pi 3B+ boards running Nexmon~\cite{nexmon} for monitor-mode access to raw IEEE~802.11 frames. To enable simultaneous transmissions, carrier sensing and random backoff are disabled, and a lightweight scheduler in the AR9271 firmware releases frames at absolute timestamps based on the local timing synchronization function (TSF) timer, with a median timing error below \SI{2}{\micro\second}.

Results show that FM4WiFi on average outperforms simple baseline schedulers: single-transmission round-robin and simultaneous transmissions by all APs. H-MAB gives higher throughput, but only after convergence. Altogether, these results confirm that FM4WiFi can be used in practical deployments.

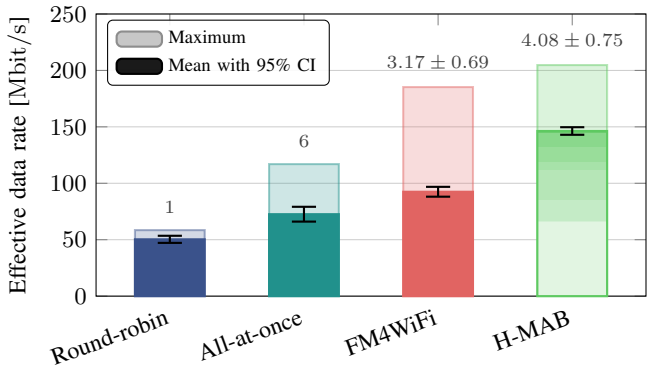
\begin{figure}
    \centering
    \begin{tikzpicture}
\colorlet{crr}{palA}
\colorlet{cat}{palB2}
\colorlet{crnd}{palG}
\colorlet{cfm}{palF}
\definecolor{chmab10}{RGB}{226,245,226}
\definecolor{chmab20}{RGB}{211,240,212}
\definecolor{chmab30}{RGB}{196,235,198}
\definecolor{chmab40}{RGB}{182,230,183}
\definecolor{chmab50}{RGB}{167,226,169}
\definecolor{chmab60}{RGB}{153,221,154}
\definecolor{chmab70}{RGB}{138,216,140}
\definecolor{chmab80}{RGB}{123,211,126}
\definecolor{chmab90}{RGB}{109,206,111}
\colorlet{chmab100}{palD}
\begin{axis}[
  width=\columnwidth,
  height=0.6\columnwidth,
  ylabel={Effective data rate [\si{\mega\bit\per\second}]},
  ymin=0, ymax=250,
  xmin=0.45, xmax=4.55,
  xtick={1,2,3,4},
  xticklabels={Round-robin, All-at-once, FM4WiFi, H-MAB},
  xticklabel style={font=\small, rotate=20, anchor=north east},
  ytick distance=50,
  ymajorgrids=true,
  xmajorgrids=false,
  major grid style={line width=0.2pt, draw=gray!40},
  tick label style={font=\small},
  label style={font=\small},
  legend style={at={(0.02,0.98)}, anchor=north west, draw=black, fill=white,
    fill opacity=0.9, text opacity=1, font=\scriptsize, rounded corners=2pt},
  legend cell align=left,
  clip=false,
]
\pgfmathsetmacro{\bw}{0.26}      

\fill[crr, fill opacity=0.22] (axis cs:{1-\bw},0) rectangle (axis cs:{1+\bw},58.500);
\draw[crr, thick, opacity=0.45] (axis cs:{1-\bw},0) rectangle (axis cs:{1+\bw},58.500);
\fill[crr] (axis cs:{1-\bw},0) rectangle (axis cs:{1+\bw},50.336);
\draw[crr, thick] (axis cs:{1-\bw},0) rectangle (axis cs:{1+\bw},50.336);
\draw[black, thick] (axis cs:1,47.183) -- (axis cs:1,53.489);
\draw[black, thick] (axis cs:{1-0.09},53.489) -- (axis cs:{1+0.09},53.489);
\draw[black, thick] (axis cs:{1-0.09},47.183) -- (axis cs:{1+0.09},47.183);
\node[font=\scriptsize, anchor=south, text=black!75] at (axis cs:1,65.375) {$1$};

\fill[cat, fill opacity=0.22] (axis cs:{2-\bw},0) rectangle (axis cs:{2+\bw},117.000);
\draw[cat, thick, opacity=0.45] (axis cs:{2-\bw},0) rectangle (axis cs:{2+\bw},117.000);
\fill[cat] (axis cs:{2-\bw},0) rectangle (axis cs:{2+\bw},72.606);
\draw[cat, thick] (axis cs:{2-\bw},0) rectangle (axis cs:{2+\bw},72.606);
\draw[black, thick] (axis cs:2,66.043) -- (axis cs:2,79.169);
\draw[black, thick] (axis cs:{2-0.09},79.169) -- (axis cs:{2+0.09},79.169);
\draw[black, thick] (axis cs:{2-0.09},66.043) -- (axis cs:{2+0.09},66.043);
\node[font=\scriptsize, anchor=south, text=black!75] at (axis cs:2,123.875) {$6$};


\fill[cfm, fill opacity=0.22] (axis cs:{3-\bw},0) rectangle (axis cs:{3+\bw},185.252);
\draw[cfm, thick, opacity=0.45] (axis cs:{3-\bw},0) rectangle (axis cs:{3+\bw},185.252);
\fill[cfm] (axis cs:{3-\bw},0) rectangle (axis cs:{3+\bw},92.487);
\draw[cfm, thick] (axis cs:{3-\bw},0) rectangle (axis cs:{3+\bw},92.487);
\draw[black, thick] (axis cs:3,88.121) -- (axis cs:3,96.853);
\draw[black, thick] (axis cs:{3-0.09},96.853) -- (axis cs:{3+0.09},96.853);
\draw[black, thick] (axis cs:{3-0.09},88.121) -- (axis cs:{3+0.09},88.121);
\node[font=\scriptsize, anchor=south, text=black!75] at (axis cs:3,192.127) {$3.17 \pm 0.69$};

\fill[chmab100, fill opacity=0.22] (axis cs:{4-\bw},0) rectangle (axis cs:{4+\bw},204.750);
\draw[chmab100, thick, opacity=0.45] (axis cs:{4-\bw},0) rectangle (axis cs:{4+\bw},204.750);
\fill[chmab100] (axis cs:{4-\bw},0) rectangle (axis cs:{4+\bw},146.324);
\fill[chmab90] (axis cs:{4-\bw},0) rectangle (axis cs:{4+\bw},145.649);
\fill[chmab80] (axis cs:{4-\bw},0) rectangle (axis cs:{4+\bw},144.488);
\fill[chmab70] (axis cs:{4-\bw},0) rectangle (axis cs:{4+\bw},144.576);
\fill[chmab60] (axis cs:{4-\bw},0) rectangle (axis cs:{4+\bw},132.149);
\fill[chmab50] (axis cs:{4-\bw},0) rectangle (axis cs:{4+\bw},118.282);
\fill[chmab40] (axis cs:{4-\bw},0) rectangle (axis cs:{4+\bw},111.234);
\fill[chmab30] (axis cs:{4-\bw},0) rectangle (axis cs:{4+\bw},84.691);
\fill[chmab20] (axis cs:{4-\bw},0) rectangle (axis cs:{4+\bw},65.958);
\fill[chmab10] (axis cs:{4-\bw},0) rectangle (axis cs:{4+\bw},65.724);
\draw[chmab100, thick] (axis cs:{4-\bw},0) rectangle (axis cs:{4+\bw},146.324);
\draw[black, thick] (axis cs:4,142.970) -- (axis cs:4,149.678);
\draw[black, thick] (axis cs:{4-0.09},149.678) -- (axis cs:{4+0.09},149.678);
\draw[black, thick] (axis cs:{4-0.09},142.970) -- (axis cs:{4+0.09},142.970);
\node[font=\scriptsize, anchor=south, text=black!75] at (axis cs:4,211.625) {$4.08 \pm 0.75$};

\addlegendimage{area legend, fill=black, fill opacity=0.22, draw=black, draw opacity=0.45}
\addlegendentry{Maximum}
\addlegendimage{area legend, fill=black, draw=black}
\addlegendentry{Mean with 95\% CI}
\end{axis}
\end{tikzpicture}
    \caption{Throughput of FM4WiFi and the baselines in the testbed: 20 MHz channels, fixed MCS 6. Numbers above the bars give the mean simultaneous transmissions ($\pm$ std. dev.). The gradient on the H-MAB bar marks its convergence period. The trends match the 2x3 AP simulation results in Fig.~\ref{fig:static_residential}.}
    \label{fig:experiments}
\end{figure}

\section{Discussion}

FM4WiFi combines features that no previous approach has, namely scalability, speed, and topology independence, but in its current form it remains a starting point. We discuss the key trade-offs and open directions below.

Because learning is performed offline, FM4WiFi cannot fine-tune to a specific deployment over time. Online methods such as \ac{H-MAB} learn the exact interference structure of a fixed topology, which is why they can outperform FM4WiFi on small scales. The offline approach is most advantageous when deployment diversity is high; a static long-term deployment could benefit from continual online adaptation on top of the pretrained model.

FM4WiFi primarily optimizes the aggregate effective data rate rather than per-station fairness. \ac{Co-SR} inherently favors stations selected for concurrent transmissions, so some stations may experience longer access delays. The current top-$k$ selection partially mitigates this, but a richer strategy maximizing coverage diversity across the sample set could better balance throughput, fairness, and latency.

The proposed deployment has two hard requirements: a central controller with wired backhaul to all \acp{AP}, and a hardware accelerator for sub-second inference. Furthermore, as a stochastic method, FM4WiFi provides no deterministic throughput or latency guarantees. In particular, individual samples may be suboptimal, and the surrogate predictor adds approximation error. Deployments with hard quality of service requirements would need additional safeguards. Finally, because the surrogate is trained on FM-generated configurations, it may not generalize well to configurations outside the FM model's support; retraining on a broader simulated distribution is a natural mitigation. Due to the absence of baselines with joint \ac{MCS} selection, the non-oracle results of FM4WiFi lack a direct point of comparison and require further validation as such baselines emerge.

Beyond addressing these limitations, future work includes extending FM4WiFi to uplink \ac{Co-SR} (not planned for Wi-Fi~8, yet easy to add in our method) and mixed-traffic scenarios, exploring continual adaptation to specific deployments, and systematically quantifying how approximation errors compound across the autoencoder, \ac{FM}, and surrogate stages.

\section{Conclusions}

The complexity of \ac{Co-SR} scheduling is a fundamental barrier to coordinated Wi-Fi at scale: existing heuristic, solver-based, and \ac{RL}-driven approaches all share one failure mode, their cost or convergence time grows with network size. We address this by reframing \ac{Co-SR} scheduling as a conditional generation problem. FM4WiFi learns the distribution of high‑quality \ac{Co-SR} configurations offline and, conditioned on the current network state, samples from it in under one second regardless of network size. This shift from search to generation unlocks Co‑SR scalability: FM4WiFi handles 30+~\acp{AP} with sub‑second inference and jointly optimizes group formation, station selection, \ac{MCS}, and transmission power. The method is both fast and effective, matching or exceeding baselines at medium and large scales and maintaining high throughput under continuous topology changes.
Crucially, this is not a simulation-only result: on a six-AP testbed built from commercial off-the-shelf hardware, FM4WiFi operated zero-shot on a previously unseen topology and outperformed the random, round-robin, and all-AP baselines.
\apxonly{ An ablation study (Appendix~\ref{sec:ablation}) confirms that each component contributes to FM4WiFi's performance.}
More broadly, for MAC-layer problems whose action space is too large to enumerate and whose environments are too diverse for per-deployment learning, our results suggest that offline generative models, following a learn-once, deploy-anywhere paradigm, are a promising way forward.

\bstctlcite{IEEEexample:BSTcontrol}
\bibliographystyle{IEEEtran}
\bibliography{bibliography}

@IEEEtranBSTCTL{IEEEexample:BSTcontrol,
  CTLuse_forced_etal       = "yes",
  CTLmax_names_forced_etal = "5",
  CTLnames_show_etal       = "1",
}

@String{Computing = "Computing" }

@String{Computer = "{IEEE} Computer" }

@online{bishop1994mdn,
  author       = {Bishop, Christopher M.},
  title        = {Mixture Density Networks},
  institution  = {Aston University},
  year         = {1994},
  url          = {https://research.aston.ac.uk/en/publications/mixture-density-networks/},
  urldate      = {2026-02-10},
  type         = {Technical Report},
}

@article{battaglia2018relational,
  author       = {Peter W. Battaglia and
                  Jessica B. Hamrick and
                  Victor Bapst and
                  Alvaro Sanchez{-}Gonzalez and
                  Vin{\'{\i}}cius Flores Zambaldi and
                  Mateusz Malinowski and
                  Andrea Tacchetti and
                  David Raposo and
                  Adam Santoro and
                  Ryan Faulkner and
                  {\c{C}}aglar G{\"{u}}l{\c{c}}ehre and
                  H. Francis Song and
                  Andrew J. Ballard and
                  Justin Gilmer and
                  George E. Dahl and
                  Ashish Vaswani and
                  Kelsey R. Allen and
                  Charles Nash and
                  Victoria Langston and
                  Chris Dyer and
                  Nicolas Heess and
                  Daan Wierstra and
                  Pushmeet Kohli and
                  Matthew M. Botvinick and
                  Oriol Vinyals and
                  Yujia Li and
                  Razvan Pascanu},
  title        = {Relational inductive biases, deep learning, and graph networks},
  journal      = {CoRR},
  volume       = {abs/1806.01261},
  year         = {2018},
  eprinttype    = {arXiv},
  eprint       = {1806.01261},
  bibsource    = {dblp computer science bibliography, https://dblp.org}
}

@misc{lipman2023flow,
      title={Flow Matching for Generative Modeling}, 
      author={Yaron Lipman and Ricky T. Q. Chen and Heli Ben-Hamu and Maximilian Nickel and Matt Le},
      year={2023},
      eprint={2210.02747},
      archivePrefix={arXiv},
      primaryClass={cs.LG}, 
}

@misc{kingma2014auto,
      title={{Auto-Encoding Variational Bayes}}, 
      author={Diederik P Kingma and Max Welling},
      year={2014},
      eprint={1312.6114},
      archivePrefix={arXiv},
      primaryClass={stat.ML}
}

@article{hendrycks2016bridging,
  author       = {Dan Hendrycks and
                  Kevin Gimpel},
  title        = {Bridging Nonlinearities and Stochastic Regularizers with Gaussian
                  Error Linear Units},
  journal      = {CoRR},
  volume       = {abs/1606.08415},
  year         = {2016},
  eprinttype    = {arXiv},
  eprint       = {1606.08415},
  bibsource    = {dblp computer science bibliography, https://dblp.org}
}

@inproceedings{vaswani2017attention,
 author = {Vaswani, Ashish and Shazeer, Noam and Parmar, Niki and Uszkoreit, Jakob and Jones, Llion and Gomez, Aidan N and Kaiser, \L ukasz and Polosukhin, Illia},
 booktitle = {Advances in Neural Information Processing Systems},
 editor = {I. Guyon and U. Von Luxburg and S. Bengio and H. Wallach and R. Fergus and S. Vishwanathan and R. Garnett},
 pages = {},
 publisher = {Curran Associates, Inc.},
 title = {Attention is All you Need},
 url = {https://proceedings.neurips.cc/paper_files/paper/2017/file/3f5ee243547dee91fbd053c1c4a845aa-Paper.pdf},
 volume = {30},
 year = {2017}
}

@inproceedings{ho2020denoising,
 author = {Ho, Jonathan and Jain, Ajay and Abbeel, Pieter},
 booktitle = {Advances in Neural Information Processing Systems},
 editor = {H. Larochelle and M. Ranzato and R. Hadsell and M.F. Balcan and H. Lin},
 pages = {6840--6851},
 publisher = {Curran Associates, Inc.},
 title = {Denoising Diffusion Probabilistic Models},
 url = {https://proceedings.neurips.cc/paper_files/paper/2020/file/4c5bcfec8584af0d967f1ab10179ca4b-Paper.pdf},
 volume = {33},
 year = {2020}
}

@ARTICLE{wojnar2025csr,
  author={Wojnar, Maksymilian and Ciężobka, Wojciech and Tomaszewski, Artur and Chołda, Piotr and Rusek, Krzysztof and Kosek-Szott, Katarzyna and Haxhibeqiri, Jetmir and Hoebeke, Jeroen and Bellalta, Boris and Zubow, Anatolij and Dressler, Falko and Szott, Szymon},
  journal={IEEE Journal on Selected Areas in Communications}, 
  title={{Coordinated Spatial Reuse Scheduling With Machine Learning in IEEE 802.11 MAPC Networks}}, 
  year={2025},
  volume={43},
  number={11},
  pages={3666-3682},
  doi={10.1109/JSAC.2025.3584555}}

@ARTICLE{wojnar2025ieee,
  author={Wojnar, Maksymilian and Ciezobka, Wojciech and Kosek-Szott, Katarzyna and Rusek, Krzysztof and Szott, Szymon and Nunez, David and Bellalta, Boris},
  journal={IEEE Communications Letters}, 
  title={{IEEE 802.11bn Multi-AP Coordinated Spatial Reuse With Hierarchical Multi-Armed Bandits}}, 
  year={2025},
  volume={29},
  number={3},
  pages={428-432},
  doi={10.1109/LCOMM.2024.3521079}}

@software{jax2018github,
  author = {James Bradbury and Roy Frostig and Peter Hawkins and Matthew James Johnson and Chris Leary and Dougal Maclaurin and George Necula and Adam Paszke and Jake Vander{P}las and Skye Wanderman-{M}ilne and Qiao Zhang},
  title = {{JAX}: composable transformations of {P}ython+{N}um{P}y programs},
  url = {http://github.com/jax-ml/jax},
  version = {0.8.0},
  year = {2018},
}

@software{deepmind2020jax,
  title = {The {D}eep{M}ind {JAX} {E}cosystem},
  author = {DeepMind and Babuschkin, Igor and Baumli, Kate and Bell, Alison and Bhupatiraju, Surya and Bruce, Jake and Buchlovsky, Peter and Budden, David and Cai, Trevor and Clark, Aidan and Danihelka, Ivo and Dedieu, Antoine and Fantacci, Claudio and Godwin, Jonathan and Jones, Chris and Hemsley, Ross and Hennigan, Tom and Hessel, Matteo and Hou, Shaobo and Kapturowski, Steven and Keck, Thomas and Kemaev, Iurii and King, Michael and Kunesch, Markus and Martens, Lena and Merzic, Hamza and Mikulik, Vladimir and Norman, Tamara and Papamakarios, George and Quan, John and Ring, Roman and Ruiz, Francisco and Sanchez, Alvaro and Sartran, Laurent and Schneider, Rosalia and Sezener, Eren and Spencer, Stephen and Srinivasan, Srivatsan and Stanojevi\'{c}, Milo\v{s} and Stokowiec, Wojciech and Wang, Luyu and Zhou, Guangyao and Viola, Fabio},
  url = {http://github.com/google-deepmind},
  year = {2020},
}

@software{flax2020github,
  author = {Jonathan Heek and Anselm Levskaya and Avital Oliver and Marvin Ritter and Bertrand Rondepierre and Andreas Steiner and Marc van {Z}ee},
  title = {{F}lax: A neural network library and ecosystem for {JAX}},
  url = {http://github.com/google/flax},
  version = {0.12.0},
  year = {2024},
}

@software{jraph2020github,
  author = {Jonathan Godwin* and Thomas Keck* and Peter Battaglia and Victor Bapst and Thomas Kipf and Yujia Li and Kimberly Stachenfeld and Petar Veli\v{c}kovi\'{c} and Alvaro Sanchez-Gonzalez},
  title = {{J}raph: {A} library for graph neural networks in jax.},
  url = {http://github.com/deepmind/jraph},
  version = {0.0.6.dev0},
  year = {2020},
}

@software{yadan2019hydra,
  author =       {Omry Yadan},
  title =        {{Hydra - A framework for elegantly configuring complex applications}},
  howpublished = {Github},
  year =         {2019},
  url =          {https://github.com/facebookresearch/hydra}
}

@software{wandb,
title = {Experiment Tracking with Weights and Biases},
year = {2020},
url={https://www.wandb.com/},
author = {Biewald, Lukas},
}

@software{grain2023github,
  author = {Marvin Ritter and Ihor Indyk and Aayush Singh and Andrew Audibert and Anoosha Seelam and Camelia Hanes and Eric Lau and Jacek Olesiak and Jiyang Kang and Xihui Wu},
  title = {{Grain} - Feeding JAX Models},
  url = {http://github.com/google/grain},
  version = {0.2.12},
  year = {2023},
}

@software{orbax2023github,
  author = {The Orbax Authors},
  title = {Orbax},
  url = {http://github.com/google/orbax},
  version = {0.11.26},
  year = {2023},
}

@misc{tgax,
    author      = {Merlin, Simone and others},
    title       = {{TGax Simulation Scenarios}},
    note      = {doc.: IEEE 802.11-14/0980r16},
    year        = 2015,
    month       = Nov,
    url = {https://mentor.ieee.org/802.11/dcn/14/11-14-0980-16-00ax-simulation-scenarios.docx}
}

@misc{loshchilov2017decoupled,
  title={Decoupled Weight Decay Regularization},
  author={Ilya Loshchilov and Frank Hutter},
  booktitle={International Conference on Learning Representations},
  eprint={1711.05101},
  year={20174}
}

@inproceedings{
orvieto2025in,
title={In Search of Adam{\textquoteright}s Secret Sauce},
author={Antonio Orvieto and Robert M. Gower},
booktitle={The Thirty-ninth Annual Conference on Neural Information Processing Systems},
year={2025},
url={https://openreview.net/forum?id=CH72XyZs4y}
}

@ARTICLE{liang2025gdsg,
  author={Liang, Ruihuai and Yang, Bo and Chen, Pengyu and Cao, Xuelin and Yu, Zhiwen and Debbah, Mérouane and Niyato, Dusit and Poor, H. Vincent and Yuen, Chau},
  journal={IEEE Transactions on Mobile Computing}, 
  title={{GDSG: Graph Diffusion-Based Solution Generator for Optimization Problems in MEC Networks}}, 
  year={2025},
  volume={24},
  number={10},
  pages={10264-10277},
  doi={10.1109/TMC.2025.3568248}}

@inproceedings{gilmer2017neural, author = {Gilmer, Justin and Schoenholz, Samuel S. and Riley, Patrick F. and Vinyals, Oriol and Dahl, George E.}, title = {Neural message passing for Quantum chemistry}, year = {2017}, publisher = {JMLR.org}, booktitle = {Proceedings of the 34th International Conference on Machine Learning - Volume 70}, pages = {1263--1272}, numpages = {10}, location = {Sydney, NSW, Australia}, series = {ICML'17} }

@INPROCEEDINGS{wilhelmi2023throughput,
  author={Wilhelmi, Francesc and Galati-Giordano, Lorenzo and Geraci, Giovanni and Bellalta, Boris and Fontanesi, Gianluca and Nuñez, David},
  booktitle={2023 IEEE Conference on Standards for Communications and Networking (CSCN)}, 
  title={Throughput Analysis of IEEE 802.11bn Coordinated Spatial Reuse}, 
  year={2023},
  volume={},
  number={},
  pages={401-407},
  doi={10.1109/CSCN60443.2023.10453190}}

@INPROCEEDINGS{haxhibeqiri2024coordinated,
  author={Haxhibeqiri, Jetmir and Jiao, Xianjun and Shen, Xiaoman and Pan, Chun and Jiang, Xingfeng and Hoebeke, Jeroen and Moerman, Ingrid},
  booktitle={2024 IEEE 20th International Conference on Factory Communication Systems (WFCS)}, 
  title={{Coordinated Spatial Reuse for WiFi Networks: A Centralized Approach}}, 
  year={2024},
  volume={},
  number={},
  pages={1-8},
  doi={10.1109/WFCS60972.2024.10540785}}

@INPROCEEDINGS{nunez2023group,
  author={Nunez, David and Smith, Malcom and Bellalta, Boris},
  booktitle={2023 21st Mediterranean Communication and Computer Networking Conference (MedComNet)}, 
  title={{Multi-AP Coordinated Spatial Reuse for Wi-Fi 8: Group Creation and Scheduling}}, 
  year={2023},
  volume={},
  number={},
  pages={203-208},
  doi={10.1109/MedComNet58619.2023.10168857}}

@INPROCEEDINGS{nunez2022txop,
  author={Nunez, D. and Wilhelmi, F. and Avallone, S. and Smith, M. and Bellalta, B.},
  booktitle={2022 IEEE 19th Annual Consumer Communications \& Networking Conference (CCNC)}, 
  title={{TXOP sharing with Coordinated Spatial Reuse in Multi-AP Cooperative IEEE 802.11be WLANs}}, 
  year={2022},
  volume={},
  number={},
  pages={864-870},
  doi={10.1109/CCNC49033.2022.9700500}}

@misc{nunez2025deep,
      title={{Deep Reinforcement Learning-Based Scheduling for Wi-Fi Multi-Access Point Coordination}}, 
      author={David Nunez and Francesc Wilhelmi and Maksymilian Wojnar and Katarzyna Kosek-Szott and Szymon Szott and Boris Bellalta},
      year={2025},
      eprint={2507.19377},
      archivePrefix={arXiv},
      primaryClass={cs.NI},
}

@article{zhu2025two,
title = {{Two enhanced schemes for coordinated spatial reuse in IEEE 802.11be: Adaptive and distributed approaches}},
journal = {Computer Networks},
volume = {258},
pages = {111060},
year = {2025},
issn = {1389-1286},
doi = {https://doi.org/10.1016/j.comnet.2025.111060},
url = {https://www.sciencedirect.com/science/article/pii/S1389128625000283},
author = {Deqing Zhu and Lidong Wang and Genmei Pan and Shenji Luan},
}

@ARTICLE{celik2024dawn,
  author={Celik, Abdulkadir and Eltawil, Ahmed M.},
  journal={IEEE Open Journal of the Communications Society}, 
  title={At the Dawn of Generative AI Era: A Tutorial-cum-Survey on New Frontiers in 6G Wireless Intelligence}, 
  year={2024},
  volume={5},
  number={},
  pages={2433-2489},
  doi={10.1109/OJCOMS.2024.3362271}}

@ARTICLE{du2024enhancing,
  author={Du, Hongyang and Zhang, Ruichen and Liu, Yinqiu and Wang, Jiacheng and Lin, Yijing and Li, Zonghang and Niyato, Dusit and Kang, Jiawen and Xiong, Zehui and Cui, Shuguang and Ai, Bo and Zhou, Haibo and Kim, Dong In},
  journal={IEEE Communications Surveys \& Tutorials}, 
  title={Enhancing Deep Reinforcement Learning: A Tutorial on Generative Diffusion Models in Network Optimization}, 
  year={2024},
  volume={26},
  number={4},
  pages={2611-2646},
  doi={10.1109/COMST.2024.3400011}}

@ARTICLE{khoramnejad2025generative,
  author={Khoramnejad, Fahime and Hossain, Ekram},
  journal={IEEE Communications Surveys \& Tutorials}, 
  title={Generative AI for the Optimization of Next-Generation Wireless Networks: Basics, State-of-the-Art, and Open Challenges}, 
  year={2025},
  volume={27},
  number={6},
  pages={3483-3525},
  doi={10.1109/COMST.2025.3535554}}

@misc{liang2025diffsg,
      title={DiffSG: A Generative Solver for Network Optimization with Diffusion Model}, 
      author={Ruihuai Liang and Bo Yang and Zhiwen Yu and Bin Guo and Xuelin Cao and Mérouane Debbah and H. Vincent Poor and Chau Yuen},
      year={2025},
      eprint={2408.06701},
      archivePrefix={arXiv},
      primaryClass={cs.NI}
}

@INPROCEEDINGS{du2025deep,
  author={Du, Mingjun and Yan, Rong and Liu, Peng and Guo, Ziyang and Sun, Xinghua},
  booktitle={2025 IEEE Wireless Communications and Networking Conference (WCNC)}, 
  title={{Deep Reinforcement Learning Based Spatial Reuse for IEEE 802.11bn}}, 
  year={2025},
  volume={},
  number={},
  pages={1-6},
  doi={10.1109/WCNC61545.2025.10978433}}

@ARTICLE{jung2026coordinated,
  author={Jung, Jaewook and Lee, Gangwoo and Chung, Jong-Moon},
  journal={IEEE Transactions on Network Science and Engineering}, 
  title={{Coordinated Spatial Reuse With Shared AP Index Optimization for IEEE 802.11be WLAN Enhancement}}, 
  year={2026},
  volume={13},
  number={},
  pages={930-947},
  doi={10.1109/TNSE.2025.3588679}}

@misc{fan2025learning,
      title={Learning Multi-Access Point Coordination in Agentic AI Wi-Fi with Large Language Models}, 
      author={Yifan Fan and Le Liang and Peng Liu and Xiao Li and Ziyang Guo and Qiao Lan and Shi Jin and Wen Tong},
      year={2025},
      eprint={2511.20719},
      archivePrefix={arXiv},
      primaryClass={cs.AI}
}

@article{ning2025survey,
  title={{A Survey on IEEE 802.11bn Wi-Fi 8: Advantages of Ultra High Reliability for Next-Generation Wireless LANs}},
  author={Ning, Wangzhong and Tang, Fengxiao and Zhao, Ming and Kato, Nei},
  journal={IEEE Communications Surveys \& Tutorials},
  volume={28},
  pages={4215--4247},
  year={2025},
  publisher={IEEE}
}

@article{val2025wi,
  title={{Wi-Fi 8 unveiled: Key features, multi-AP coordination, and the role of C-TDMA}},
  author={Val, I{\~n}aki and L{\'o}pez-P{\'e}rez, David and Kijanka, Aleksandra and Schelstraete, Sigurd and Mu{\~n}oz, Luis and Arlandis, Diego and Mart{\'\i}nez, Marcos},
  journal={IEEE Communications Magazine},
  year={2025},
  publisher={IEEE}
}

@misc{802.11-24/0209r12,
  title={{Specification Framework for TGbn}},
  author={Ross Jian Yu},
  year={2025},
  url={https://mentor.ieee.org/802.11/dcn/24/11-24-0209-12-00bn-specification-framework-for-tgbn.docx},
note="Accessed on March 2, 2026"
}

@misc{802.11-26/0287r1,
  title={{LB291 CR for Co-BF and Co-SR MISC Part 2}},
  author={Jason Yuchen Guo et al.},
  year={2026},
  url={https://mentor.ieee.org/802.11/dcn/26/11-26-0287-01-00bn-lb291-cr-for-co-bf-and-co-sr-misc-part-2.docx},
note="Accessed on March 2, 2026"
}

@misc{modwifi,
  author       = {Mathy Vanhoef},
  title        = {Modwifi: Modified Wi-Fi Stack for Wireless Networking Research},
  year         = {2024},
  howpublished = {\url{https://github.com/vanhoefm/modwifi}}
}

@misc{nexmon,
  author       = {{SEEMOO Lab}},
  title        = {Nexmon: The C-based Firmware Patching Framework},
  year         = {2024},
  howpublished = {\url{https://github.com/seemoo-lab/nexmon}}
}

\ifhideappendix\else
\clearpage

\appendices

\section{Related Work}
\label{sec:related_work}

The IEEE 802.11bn task group has defined a common framework for \ac{MAPC} coordination (including \ac{Co-SR}) and procedures for discovery and negotiation~\cite{802.11-24/0209r12}. The expected adoption of \ac{MAPC} in future Wi‑Fi networks~\cite{ning2025survey} has motivated research on optimizing \ac{Co-SR}, yet existing approaches struggle to balance decision quality with strict real-time MAC constraints.

\paragraph{Analytical and heuristic models}
Early \ac{CTMC}-based studies show up to 59\% throughput gains~\cite{wilhelmi2023throughput} but rely on saturated traffic assumptions. Practical systems typically use centralized controllers driven by periodic \ac{RSSI} reports~\cite{haxhibeqiri2024coordinated, nunez2023group}, which introduce a latency loop that degrades performance under mobility. Distributed schemes using convex optimization~\cite{zhu2025two} or \ac{Co-TDMA}~\cite{nunez2022txop} reduce signaling but remain intractable at scale due to the combinatorial group-formation space.

\paragraph{RL in MAPC}
To overcome the rigidity of heuristics, \ac{RL} has been widely adopted. \Ac{MAB} formulations provide robust adaptation without a priori knowledge of the channel~\cite{wojnar2025ieee, wojnar2025csr}. Deep \ac{RL} (\ac{DRL}) approaches offer joint interference detection and power control, often using multi-agent frameworks to decentralize execution~\cite{du2025deep, jung2026coordinated}. Recent work~\cite{nunez2025deep} has demonstrated improved latency by scheduling pre-formed \ac{Co-SR} configurations. Schemes such as SAIOC~\cite{jung2026coordinated} require more than 20,000 training episodes to converge and no public implementation is available. In general, a fundamental limitation identified by surveys is that \ac{DRL} agents often exhibit high sensitivity to reward function design and lack of robustness to dynamic distribution shifts in network state~\cite{celik2024dawn, du2024enhancing}.

\paragraph{Generative AI for wireless optimization} 
Beyond \ac{RL}, recent research has explored generative \ac{AI} as a means to model the complex underlying distributions of feasible network states~\cite{celik2024dawn, khoramnejad2025generative}. Although early efforts used \acp{GAN} for channel estimation and resource allocation, surveys highlight their inherent training instability and the tendency toward mode collapse in high-dimensional wireless data~\cite{du2024enhancing}. This has led to an emerging preference for \acp{GDM}, which offer superior stability and the ability to capture multi-modal action distributions~\cite{du2024enhancing, liang2025diffsg}. Nevertheless, \acp{GDM} suffer from an iterative ``denoising bottleneck'' that restricts their application to static or slow-varying snapshots. Simultaneously, agentic \ac{AI} frameworks utilizing \acp{LLM} have been proposed for multi-\ac{AP} negotiation~\cite{fan2025learning}, although their multi-second inference latency remains orders of magnitude too slow for MAC-layer scheduling.

\section{N/A-Gated Categorical Loss Derivation}
\label{sec:na_loss_derivation}

The decoder outputs $C$ logits $\mathbf{l} \in \mathbb{R}^C$, interpreted as a mixture via
\begin{equation}
    p_{\mathsf{N/A}} = \sigma(l_C), \qquad \boldsymbol{\pi} = \mathrm{softmax}(\mathbf{l}_{1:C-1}).
    \label{eq:na_gate}
\end{equation}
The mixture assigns probability $p_{\mathsf{N/A}}$ when $y_C = 1$ (attribute is $\mathsf{N/A}$), and probability $(1 - p_{\mathsf{N/A}})\,P_{\mathrm{cat}}(\mathbf{y}_{1:C-1})$ when $y_C = 0$ (attribute takes a valid class), where $P_{\mathrm{cat}}(\mathbf{y}_{1:C-1}) = \prod_{i=1}^{C-1}\pi_i^{\,y_i}$. Using $y_C$ as an indicator to unify both cases:
\begin{equation}
    P(\mathbf{y}) = p_{\mathsf{N/A}}^{\,y_C}
    \cdot \Bigl[(1 - p_{\mathsf{N/A}})\,P_{\mathrm{cat}}(\mathbf{y}_{1:C-1})\Bigr]^{1-y_C}.
    \label{eq:mixture_likelihood}
\end{equation}
Taking the negative logarithm of~\eqref{eq:mixture_likelihood} and expanding:
\begin{align}
    -\log P(\mathbf{y})
    &= -y_C \log p_{\mathsf{N/A}}
      - (1-y_C)\log(1 - p_{\mathsf{N/A}}) \nonumber \\
    &\quad - (1-y_C)\log P_{\mathrm{cat}}(\mathbf{y}_{1:C-1}).
    \label{eq:nll_expanded}
\end{align}
Substituting $p_{\mathsf{N/A}} = \sigma(l_C)$ and $\log P_{\mathrm{cat}}(\mathbf{y}_{1:C-1}) = \sum_{i=1}^{C-1} y_i \log \pi_i$ into~\eqref{eq:nll_expanded}, each term is recognized as a standard loss:
\begin{align}
-\log P(\mathbf{y}) &=
\underbrace{
  -y_C \log \sigma(l_C)
  \;-\;
  (1-y_C)\log\!\bigl(1-\sigma(l_C)\bigr)
}_{\mathcal{L}_{\mathrm{BCE}}(l_C,\,y_C)} \nonumber \\
&\quad+\;
\underbrace{
  (1-y_C)\!\left[
    -\sum_{i=1}^{C-1} y_i \log \pi_i
  \right]
}_{(1-y_C)\,\mathcal{L}_{\mathrm{CE}}(\mathbf{l}_{1:C-1},\,\mathbf{y}_{1:C-1})}.
\end{align}
This gives the final form~\eqref{eq:mixture_loss}:
\begin{equation*}
    \mathcal{L}_{\mathrm{mix}}(\mathbf{l}, \mathbf{y}) = \mathcal{L}_{\mathrm{BCE}}(l_C, y_C) + (1 - y_C)\,\mathcal{L}_{\mathrm{CE}}(\mathbf{l}_{1:C-1}, \mathbf{y}_{1:C-1}).
\end{equation*}

\section{Architecture and Training Details}
\label{sec:architecture_diagrams}

Figures~\ref{fig:autoencoder_diagram}--\ref{fig:surrogate_diagram} illustrate the training-time operation of the three pipeline components. The inference pipeline, which combines these components, is shown in Fig.~\ref{fig:inference_pipeline} in the main text. Table~\ref{tab:optimizer_hparams} summarizes the architecture and training hyperparameters.

The entire pipeline is implemented in JAX~\cite{jax2018github} with \ac{GNN} primitives provided by Jraph~\cite{jraph2020github} and neural network modules defined using Flax~\cite{flax2020github}.
Optimization is handled by Optax~\cite{deepmind2020jax}, checkpointing by Orbax~\cite{orbax2023github}, and data loading by Grain~\cite{grain2023github}.
Hydra~\cite{yadan2019hydra} manages configuration files and the Weights \& Biases~\cite{wandb} platform tracks training experiments.

We use \texttt{mapc\_sim}~\cite{wojnar2025ieee}, an open-source JAX-based IEEE 802.11 simulator that models \ac{MAPC} operation, including \ac{Co-SR} frame exchanges. It computes per-link data rates based on \ac{SINR}, taking into account interference from concurrent transmissions.

\begin{table*}[t]
\caption{AdamW optimizer and training hyperparameters for each model family. Within each family, all settings except the learning rate are shared across sizes; the learning rate is tuned independently per size.}
\label{tab:optimizer_hparams}
\centering
\small
\begin{tabular}{c|cccccccccc}
\toprule
Model Family & $\beta_1$ & $\beta_2$ & Weight Decay & Batch Size & Steps & Grad Clip & LR (Tiny) & LR (Small) & LR (Medium) & LR (Large)\\
\midrule
Autoencoder  & 0.95 & 0.95 & 0.01 & 128 & 3000 & 1.0 & \num{2e-4} & \num{1e-4} & \num{5e-5} & \num{3e-5}\\
Flow matching & 0.95 & 0.95 & 0.01 & 32 & 5000 & 1.0 & \num{2e-2} & \num{3e-3} & \num{1e-3} & \num{2e-4}\\
Surrogate    & 0.9 & 0.999 & 0.01 & 32 & 5000 & 1.0 & \num{1e-3} & \num{3e-4} & \num{5e-4} & \num{5e-5}\\
\bottomrule
\end{tabular}
\end{table*}

\begin{figure}[t]
    \centering
    \includegraphics[width=\linewidth]{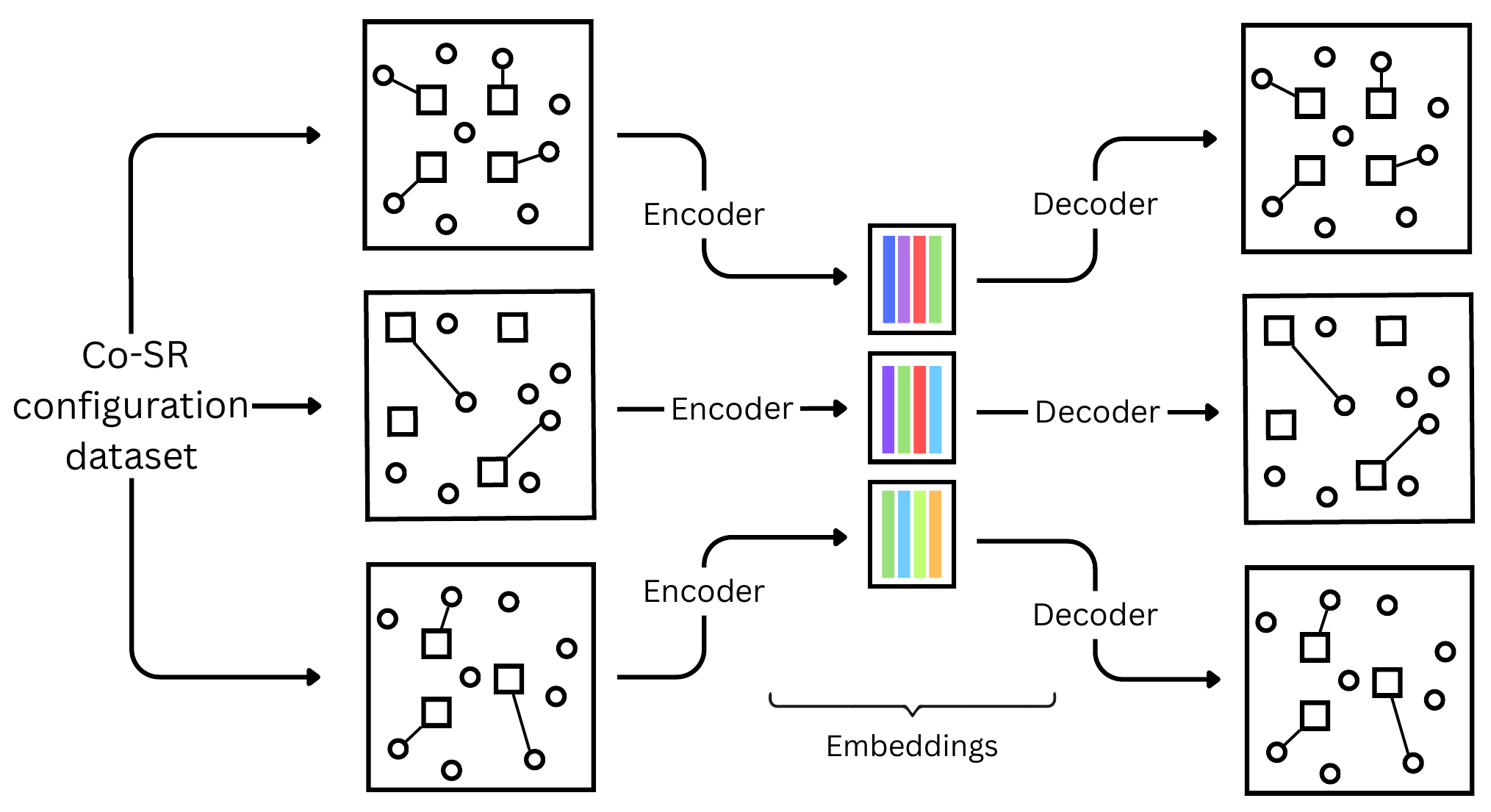}
    \caption{GNN autoencoder: the encoder maps edge attributes to per-edge latent vectors $\mathbf{z}_e$; the decoder reconstructs the original attributes from the latent graph.}
    \label{fig:autoencoder_diagram}
\end{figure}

\begin{figure}[t]
    \centering
    \includegraphics[width=\linewidth]{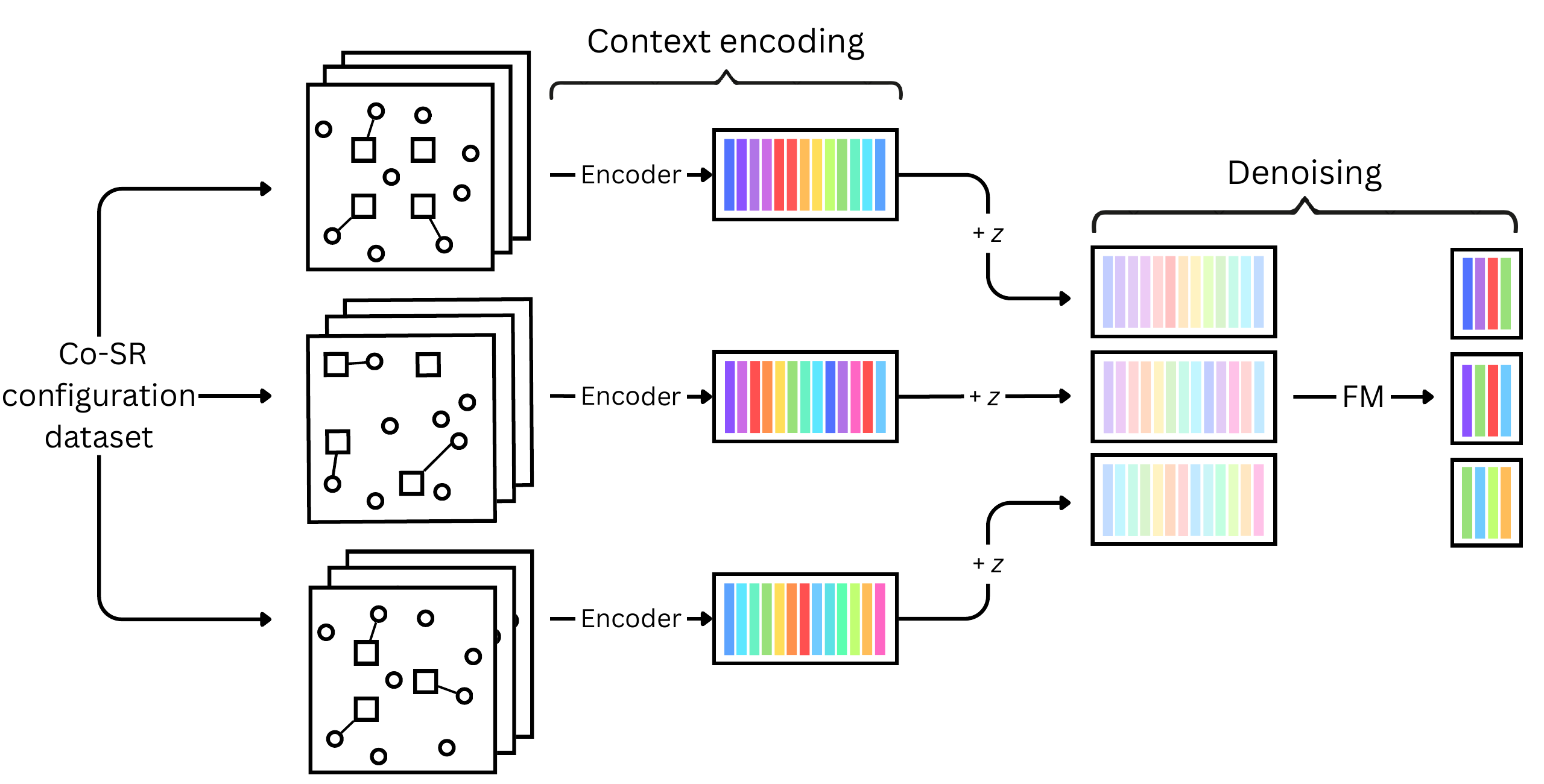}
    \caption{Flow matching training: a velocity network learns to transport noise toward the data distribution, conditioned on encoded context configurations.}
    \label{fig:fm_diagram}
\end{figure}

\begin{figure}[t]
    \centering
    \includegraphics[width=0.65\linewidth]{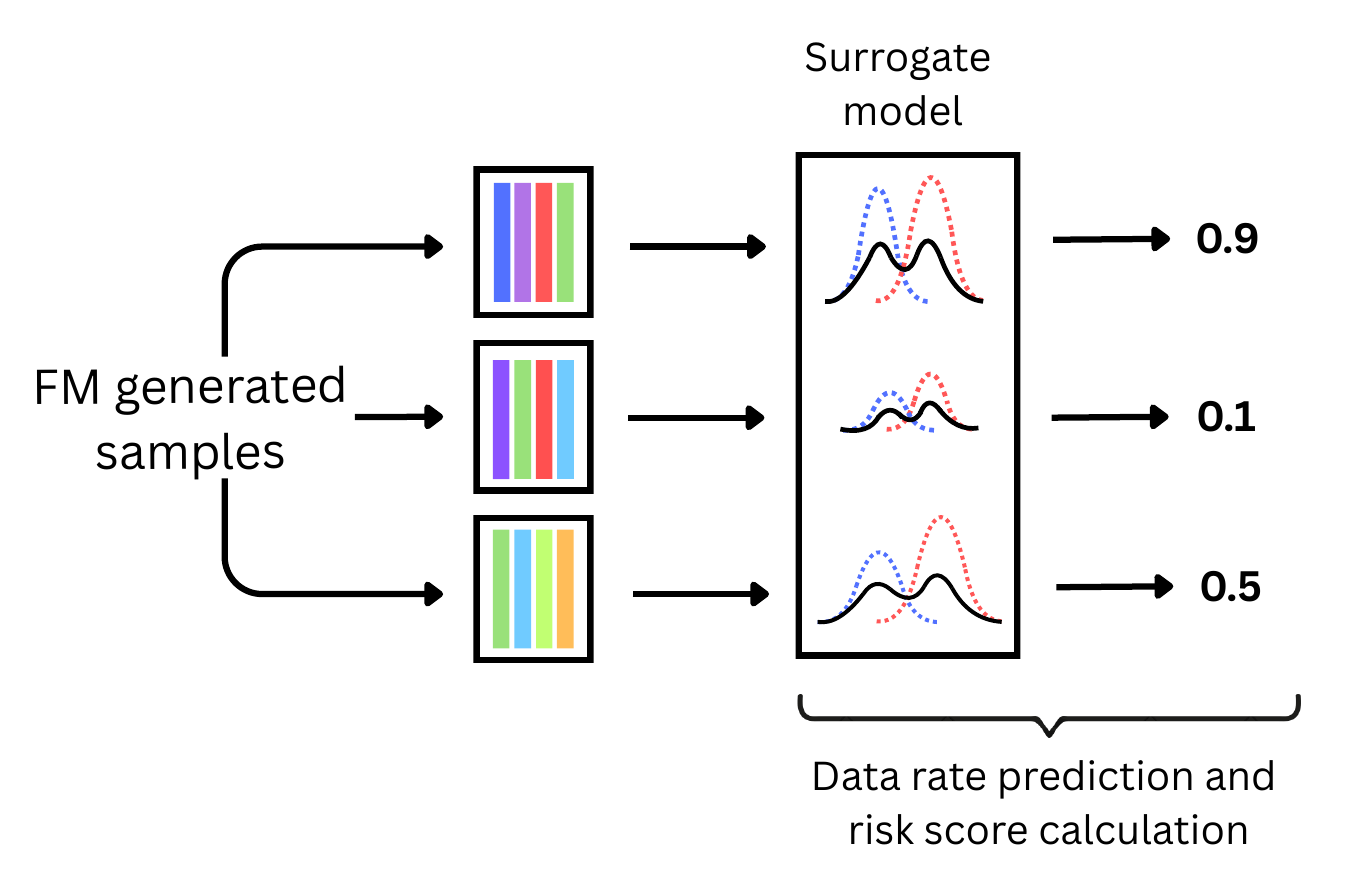}
    \caption{Surrogate data rate predictor: the model takes an encoded graph and outputs the parameters of a Gaussian mixture predicting the effective data rate.}
    \label{fig:surrogate_diagram}
\end{figure}

\subsection{More Training Details}
\label{sec:training_details}

All three FM4WiFi models (i.e., autoencoder, \ac{FM}, and surrogate) are trained using the AdamW optimizer~\cite{loshchilov2017decoupled} with gradient clipping (global norm $\leq 1.0$). The autoencoder and \ac{FM} use $(\beta_1, \beta_2) = (0.95, 0.95)$ following~\cite{orvieto2025in}; the surrogate uses the standard values $(0.9, 0.999)$ for training stability. The autoencoder and \ac{FM} models use a constant learning rate, while the surrogate model uses a cosine learning rate schedule to improve convergence in the last phase of training. Graphs are batched and padded to power-of-two sizes for efficient GPU compilation. All models use \texttt{float64} precision to ensure numerical stability. Table~\ref{tab:optimizer_hparams} (Appendix~\ref{sec:architecture_diagrams}) summarizes the architecture and training hyperparameters for each model. For performance tests, we use a server with NVIDIA GH200, \SI{118}{\giga\byte} of RAM, and Rocky Linux 9.

\subsubsection{Autoencoder}
We set the \ac{KL} divergence weight $\lambda_{\text{KL}} = 0.01$ to balance two competing goals: the \ac{FM} model assumes a Gaussian source distribution $p(\mathbf{z}) = \mathcal{N}(\mathbf{0}, \mathbf{I})$ when generating samples from noise, so the aggregate posterior should approximate this prior; at the same time, too large a $\lambda_{\text{KL}}$ risks posterior collapse, where $q_\phi(\mathbf{z} | \mathcal{G})$ ignores the input and the decoder loses reconstruction fidelity. We verify this setting in Section~\ref{sec:autoencoder_ablations}. During inference, only the mean~$\boldsymbol{\mu}_e$ is used (no sampling), ensuring deterministic encoding. Following the graph network framework, information is stored at three levels: \emph{nodes} represent individual \acp{AP} and \acp{STA} (initialized without features, they accumulate information through message passing), \emph{edges} carry per-link attributes, and a \emph{global} vector (initially empty) summarizes the entire network state.

\subsubsection{Flow Matching}
The \ac{FM} model loads the frozen encoder from a trained autoencoder checkpoint. During data preparation, each training example is augmented with $n$ context configurations (ablated in Section~\ref{sec:fm_ablations}; we find $n=1$ to be optimal) generated by running random transmissions through the simulator and encoding the resulting graphs. The time variable $t$ is sampled uniformly from $[0, 0.99)$. Both standard and \ac{EMA} parameters are tracked, with \ac{EMA} parameters used for validation and inference.

\subsubsection{Surrogate}
The surrogate is trained on configurations generated by the \ac{FM} model and evaluated in the simulator, rather than on the original baseline dataset. This ensures that the surrogate learns to score the kinds of configurations actually produced at inference time. Like the \ac{FM} model, the surrogate operates on latent representations of the network. The number of components in the mixture is fixed at $m=4$ as a practical quality--complexity trade-off (Section~\ref{sec:surrogate_ablations}), while larger mixtures can slightly reduce validation \ac{NLL}. We ensure that the scales $\boldsymbol{\sigma}$ remain positive by applying softplus activation.

\section{Ablation Studies}
\label{sec:ablation}

We perform ablation studies to isolate the impact of key modeling and optimization choices. Because the full pipeline has inter-model dependencies, each model is ablated independently and evaluated with model-specific metrics to avoid confounding effects across stages; consequently, this section does not report end-to-end metrics. We vary one factor at a time and keep other settings unchanged to make comparisons attributable to the modified component; when a new setting is found to be beneficial, we adopt it for all subsequent ablations and main experiments. For each model family (autoencoder, flow matching, surrogate) and each size (tiny, small, medium, large), we tune the learning rate. Model sizes correspond to approximately 10k (tiny), 100k (small), 1M (medium), and 10M (large) trainable parameters, excluding time embeddings. Due to the large number of configurations explored, each ablation is a single run; the main evaluation results in the preceding sections use multiple seeds with statistical analysis. All ablation experiments use a \SI{20}{\mega\hertz} channel bandwidth; while absolute data rates are lower than in the main evaluation, the relative comparisons remain qualitatively valid. In all tables, the best value per group is shown in \textbf{bold}; when a second-best result is within the rounding precision, it is also in bold.

\subsection{Autoencoder}
\label{sec:autoencoder_ablations}

\paragraph{KL regularization strength.}
As shown in Table~\ref{tab:gnn_ablation}, with $\beta=0$, the autoencoder reaches a near-perfect reconstruction but produces an unregularized latent space (high \ac{KL} divergence). In contrast, strong regularization ($\beta=1$) damages reconstruction, especially for \ac{RSSI}. In our setting, $\beta=0.01$ provides the best balance: it preserves near-perfect \ac{MCS}/transmission power accuracy\footnote{For categorical attributes, \emph{$\mathsf{N/A}$ accuracy} measures correct $\mathsf{N/A}$ vs.\ non-$\mathsf{N/A}$ classification; \emph{prediction accuracy} measures correct class prediction among non-$\mathsf{N/A}$ edges. For continuous attributes, prediction accuracy counts edges with an absolute error below~$0.1$.} while substantially reducing the \ac{KL} term, suggesting a more structured latent representation without sacrificing fidelity. From this point on, we set $\beta=0.01$ and use it for all subsequent ablation studies and experiments.

\begin{table}[t]
\caption{Effect of KL regularization strength~$\beta$ on \ac{GNN} autoencoder reconstruction accuracy and latent-space regularity. $\beta{=}0$ yields near-perfect reconstruction but an unstructured latent space; $\beta{=}0.01$ preserves high accuracy while substantially reducing KL divergence.}
\label{tab:gnn_ablation}
\centering
\footnotesize
\begin{tabular}{>{\centering}p{0.8cm}c|>{\centering}p{1.25cm}>{\centering}p{1.2cm}>{\centering}p{1.2cm}>{\centering\arraybackslash}p{1.2cm}}
\toprule
Model Size & $\beta$ & \ac{MCS} Accuracy [\%]~$\uparrow$ & TX Power Accuracy [\%]~$\uparrow$ & \ac{RSSI} Accuracy [\%]~$\uparrow$ & KL Divergence~$\downarrow$\\
\midrule
  \multirow{3}{*}{Tiny}  & 0.0  & 98.7 & \textbf{99.6} & 97.6 & 61.806\\
                         & 0.01 & 96.2 & 99.3 & 72.1 & 1.028\\
                         & 1.0  & 26.2 & 31.8 & 17.1 & 0.174\\
\specialrule{0.4pt}{2pt}{2pt}
  \multirow{3}{*}{Small} & 0.0  & \textbf{100.0} & \textbf{100.0} & 98.8 & 14.568\\
                         & 0.01 & \textbf{99.9} & \textbf{100.0} & 88.4 & 0.325\\
                         & 1.0  & 71.9 & 86.3 & 27.2 & 0.027 \\
\specialrule{0.4pt}{2pt}{2pt}
  \multirow{3}{*}{Medium} & 0.0  & \textbf{100.0} & \textbf{100.0} & \textbf{99.4} & 6.735\\
                          & 0.01 & \textbf{100.0} & \textbf{100.0} & 92.2 & 0.183\\
                          & 1.0  & 91.4 & 94.8 & 30.3 & \textbf{0.019}\\
\specialrule{0.4pt}{2pt}{2pt}
  \multirow{3}{*}{Large}  & 0.0  & \textbf{100.0} & \textbf{100.0} & \textbf{99.7} & 5.226\\
                          & 0.01 & \textbf{100.0} & \textbf{100.0} & 91.8 & 0.110\\
                          & 1.0  & 97.1 & 97.8 & 34.5 & \textbf{0.013}\\
\bottomrule
\end{tabular}
\end{table}

\paragraph{N/A-gated categorical loss.}
We compare our proposed N/A-gated categorical loss against a classic softmax baseline. Table~\ref{tab:loss_vs_softmax_top5} reports the five metrics with the most significant differences in accuracy, which occur predominantly for the tiny and small models. This concentration indicates that our proposed loss function is particularly important for preserving performance at smaller scales. The largest regression appears for \texttt{Selected (N/A)} in the small model (absolute drop of 12.87 pp.), indicating that softmax substantially degrades the link-selection decision in this setting. For the remaining metrics, the gap is comparatively small (about \SIrange{1}{2}{\percent} absolute), suggesting that softmax preserves prediction accuracy (e.g., \ac{MCS} and transmission power values) but struggles with the higher-level objectives.

\begin{table}[t]
\caption{Five largest accuracy deltas when replacing our proposed N/A-gated categorical loss (\ref{eq:mixture_loss}) with standard softmax cross-entropy (negative values indicate softmax is worse).}
\label{tab:loss_vs_softmax_top5}
\centering
\footnotesize
\begin{tabular}{>{\centering}p{1.8cm}>{\centering}p{1cm}|>{\centering}p{1.4cm}>{\centering}p{1.4cm}>{\centering\arraybackslash}p{1cm}}
\toprule
Feature & Model Size & N/A-gated Accuracy [\%] & Softmax Accuracy [\%] & Difference [\%]\\
\midrule
Selected (N/A) & Small  & 100.00 & 87.13 & -12.87\\
\ac{MCS} (pred.) & Tiny & 96.19 & 94.14 & -2.13\\
Selected (N/A) & Tiny   & 99.99 & 98.07 & -1.91\\
Active (N/A) & Medium  & 100.00 & 98.22 & -1.78\\
Transmission power (pred.) & \multirow{2}{*}{Tiny} & \multirow{2}{*}{99.25} & \multirow{2}{*}{98.07} & \multirow{2}{*}{-1.19}\\
\bottomrule
\end{tabular}
\end{table}

\paragraph{Dimensionality.}
Table~\ref{tab:latent_space} summarizes the \ac{PCA} structure of the learned latent spaces depending on the dimension $d_z \in \{6, 16, 32, 64\}$. Effective utilization grows sub-linearly with $d_z$: 95\% of variance is explained by 5 out of 6 dimensions for tiny but only 17 out of 64 for large. The variance also becomes more evenly spread: the first principal component accounts for 47\% of variance in the tiny model and only 33\% in large. These observations suggest that the Wi-Fi configuration space has a latent dimensionality of roughly 10--17, so that 32 latent dimensions are sufficient for high-fidelity encoding, while 64 dimensions add redundancy without clear benefit. Therefore, we use the medium-sized model for all experiments.

\begin{table}[t]
\caption{\ac{PCA} characterization of the learned latent space for four autoencoder sizes ($\beta{=}0.01$). The number of principal components required to explain 90\%, 95\%, and 99\% of variance grows sub-linearly with dimensionality, suggesting an intrinsic dimensionality of roughly 10--17.}
\label{tab:latent_space}
\centering
\small
\begin{tabular}{l|cccc}
\toprule
Metric & Tiny & Small & Medium & Large\\
\midrule
Latent dim & 6 & 16 & 32 & 64\\
\ac{PCA} dims (90\% var.) & 4 & 7 & 9 & 13\\
\ac{PCA} dims (95\% var.) & 5 & 8 & 11 & 17\\
\ac{PCA} dims (99\% var.) & 6 & 10 & 18 & 34\\
Top-1 \ac{PCA} explained var. & 47\% & 38\% & 34\% & 33\% \\
\bottomrule
\end{tabular}
\end{table}

\paragraph{Interpolation quality.}
To assess whether interpolated latent vectors remain realistic, we use the \ac{MMD}, a kernel-based distance between probability distributions. Fig.~\ref{fig:interpolation_mmd} reports \ac{MMD} between interpolants and the training distribution at ratios $\alpha \in \{0, 0.1, \ldots, 1\}$. All models exhibit a gentle,  midpoint peak (maximum near $\alpha \approx 0.4$), yet the absolute increase is small ($<0.01$ \ac{MMD}). Crucially, all interpolants remain far below the random noise baselines: tiny's baseline is 0.43 (lower due to the compact 6-dim space), while the baselines for other sizes are $0.73$--$0.76$ (not shown in the figure). This confirms that interpolated latent vectors stay within the learned manifold rather than drifting into unstructured regions.

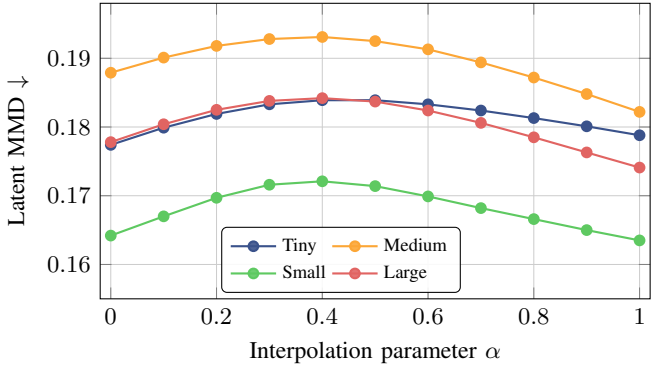
\begin{figure}[t]
\centering
\begin{tikzpicture}
\colorlet{ctiny}{palA}
\colorlet{csmall}{palD}
\colorlet{cmedium}{palH}
\colorlet{clarge}{palF}

\begin{axis}[
  width=\columnwidth,
  height=0.62\columnwidth,
  xlabel={Interpolation parameter $\alpha$},
  ylabel={Latent MMD~$\downarrow$},
  ymin=0.155, ymax=0.198,
  xmin=-0.02, xmax=1.02,
  xtick={0,0.2,0.4,0.6,0.8,1.0},
  ytick={0.16,0.17,0.18,0.19},
  grid=both,
  major grid style={line width=0.2pt, draw=gray!40},
  legend columns=2,
  legend style={at={(0.22,0.02)}, anchor=south west, draw=black, fill=white,
    fill opacity=0.9, text opacity=1, font=\scriptsize, rounded corners=2pt},
  legend cell align=left,
  tick label style={font=\small},
  label style={font=\small},
  every axis plot/.append style={thick, mark=*, mark size=1.8pt},
]
\addplot[color=ctiny] coordinates {
  (0,0.1774)(0.1,0.1799)(0.2,0.1819)(0.3,0.1833)(0.4,0.1839)
  (0.5,0.1839)(0.6,0.1833)(0.7,0.1824)(0.8,0.1813)(0.9,0.1801)(1,0.1788)};
\addlegendentry{Tiny}
\addplot[color=cmedium] coordinates {
  (0,0.1879)(0.1,0.1901)(0.2,0.1918)(0.3,0.1928)(0.4,0.1931)
  (0.5,0.1925)(0.6,0.1913)(0.7,0.1894)(0.8,0.1872)(0.9,0.1848)(1,0.1822)};
\addlegendentry{Medium}
\addplot[color=csmall] coordinates {
  (0,0.1642)(0.1,0.1670)(0.2,0.1697)(0.3,0.1716)(0.4,0.1721)
  (0.5,0.1714)(0.6,0.1699)(0.7,0.1682)(0.8,0.1666)(0.9,0.1650)(1,0.1635)};
\addlegendentry{Small}
\addplot[color=clarge] coordinates {
  (0,0.1778)(0.1,0.1804)(0.2,0.1825)(0.3,0.1838)(0.4,0.1842)
  (0.5,0.1837)(0.6,0.1824)(0.7,0.1806)(0.8,0.1785)(0.9,0.1763)(1,0.1741)};
\addlegendentry{Large}
\end{axis}
\end{tikzpicture}
\caption{\ac{MMD} between linearly interpolated latent vectors and the training distribution as a function of the interpolation ratio~$\alpha$. All models show a gentle midpoint peak at $\alpha \approx 0.4$, yet remain far below the random-noise baselines (tiny: 0.43, small: 0.73, medium: 0.76, large: 0.76), confirming that interpolants stay on the learned manifold.}
\label{fig:interpolation_mmd}
\end{figure}

\subsection{Flow Matching}
\label{sec:fm_ablations}

We ablate the key design choices of the \ac{FM} model: latent-space formulation and denoising scope, model scaling, \ac{EMA} of parameters, timestep sampling schedule, number of context observations, and \ac{ODE} solver configuration. Unless stated otherwise, ablations use the medium-sized model with $n=7$ context observations and the Euler solver with 10 integration steps. When a better configuration is identified, it is adopted for all subsequent ablations and main experiments; the final configuration is $n=1$ with 6 Euler steps.

\paragraph{Formulation.}
Table~\ref{tab:fm_comparison} compares three \ac{FM} variants: (i)~continuous \ac{FM} denoising only edge features (\emph{edges only}), (ii)~continuous \ac{FM} denoising edges, nodes, and the global vector jointly (\emph{full graph}), and (iii)~a discrete \ac{FM} operating on raw integer attribute indices rather than latent vectors. Continuous edges-only \ac{FM} achieves the highest \ac{MCS} accuracy (up to \SI{25.2}{\percent} for the large model) and transmission power accuracy (up to \SI{26.3}{\percent} for the large model), while maintaining comparable selected and success accuracy. Denoising the full graph yields no benefit, as node and global features carry no scheduling information beyond what the graph structure already encodes. The discrete formulation performs substantially worse on scheduling-decision metrics (i.e., \ac{MCS} and transmission power). This gap is expected for two reasons: first, continuous \ac{FM} relies on continuous interpolation paths (\ref{eq:interpolation}), and the latent space of the autoencoder provides a smooth manifold where such paths are well-defined, unlike the raw combinatorial attribute space; second, the continuous variants receive a pre-trained graph representation from the autoencoder, whereas the discrete model must learn its own internal representation from scratch within the same parameter and training budget. We therefore adopt continuous edges-only \ac{FM} for all subsequent experiments.

\begin{table*}[t]
\caption{Per-attribute prediction accuracy of generated \ac{Co-SR} configurations for three \ac{FM} formulations across four model sizes. Continuous edges-only achieves the best scheduling-decision accuracy (\ac{MCS}, TX~power), while the discrete variant performs substantially worse despite comparable structural accuracy (selected, success).}
\label{tab:fm_comparison}
\centering
\small
\begin{tabular}{cc|cccc}
\toprule
Size & Type & \ac{MCS} Acc. [\%]~$\uparrow$ & TX Power Acc. [\%]~$\uparrow$ & Selected Acc. [\%]~$\uparrow$ & Success Acc [\%]~$\uparrow$\\
\midrule
  \multirow{3}{*}{Tiny}    & Cont. \ac{FM} (edges only) & 21.2 & 25.7 & \textbf{85.9} & \textbf{92.4}\\
                           & Cont. \ac{FM}    & 21.1 & 25.8 & 82.3 & 89.4\\
                           & Discrete \ac{FM} & 3.3  & 4.7  & 82.2 & 89.7\\
\specialrule{0.4pt}{2pt}{2pt}
  \multirow{3}{*}{Small}   & Cont. \ac{FM} (edges only) & 22.8 & 24.8 & \textbf{83.1} & \textbf{90.5}\\
                           & Cont. \ac{FM}    & 22.4 & 25.5 & 81.1 & 88.8\\
                           & Discrete \ac{FM} & 5.7  & 6.7  & 81.0 & 88.8\\
\specialrule{0.4pt}{2pt}{2pt}
  \multirow{3}{*}{Medium}  & Cont. \ac{FM} (edges only) & \textbf{25.0} & \textbf{26.0} & 80.7 & 88.7\\
                           & Cont. \ac{FM}    & 24.0 & 25.9 & 79.9 & 88.1\\
                           & Discrete \ac{FM} & 7.6  & 7.8  & 80.7 & 88.7\\
\specialrule{0.4pt}{2pt}{2pt}
  \multirow{3}{*}{Large}   & Cont. \ac{FM} (edges only) & \textbf{25.2} & \textbf{26.3} & 80.6 & 88.8\\
                           & Cont. \ac{FM}    & 23.6 & 25.8 & 80.2 & 88.2\\
                           & Discrete \ac{FM} & 7.3  & 7.2  & 81.1 & 89.0\\
\bottomrule
\end{tabular}
\end{table*}

\paragraph{Model scaling.}
Fig.~\ref{fig:fm_ema} (solid lines) shows the validation loss and \ac{MMD} as a function of the model size. Both metrics improve steadily from tiny to medium, but the large model does not improve over medium (loss: $0.098$ vs.\ $0.087$; \ac{MMD}: $0.076$ vs.\ $0.071$). As with the surrogate (Section~\ref{sec:surrogate_ablations}), all sizes are trained for the same number of optimization steps (Table~\ref{tab:optimizer_hparams}), so the large model likely requires a longer training budget to fully converge. Since medium achieves the best quality--cost tradeoff, we use it for all subsequent \ac{FM} experiments.

\paragraph{Exponential moving average.}
Fig.~\ref{fig:fm_ema} (dashed lines) shows that \ac{EMA} with decay $\beta=0.999$ slightly improves \ac{MMD} for medium and large models, with the largest gain at medium size where \ac{MMD} is halved from $0.071$ to $0.035$. For tiny and small models, the effect is negligible. Since \ac{EMA} adds negligible computational overhead but improves distributional quality at the scales we deploy, we use \ac{EMA} parameters for all validation and inference.

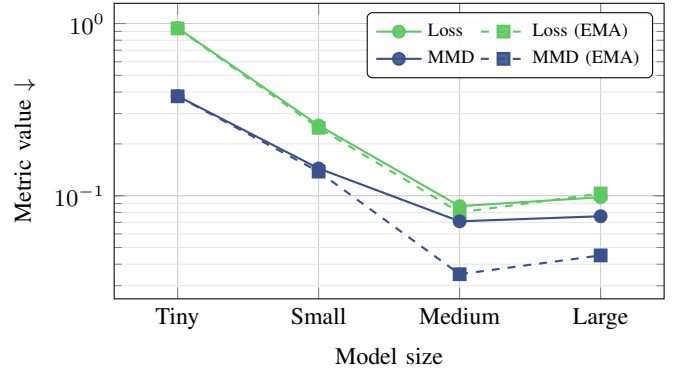
\begin{figure}[t]
\centering
\begin{tikzpicture}
\colorlet{closs}{palD}
\colorlet{cmmd}{palA}

\begin{axis}[
  width=\columnwidth,
  height=0.62\columnwidth,
  xlabel={Model size},
  ylabel={Metric value~$\downarrow$},
  ymode=log,
  symbolic x coords={Tiny,Small,Medium,Large},
  xtick=data,
  grid=both,
  major grid style={line width=0.2pt, draw=gray!40},
  minor grid style={line width=0.1pt, draw=gray!20},
  enlarge x limits=0.15,
  clip=false,
  legend columns=2,
  legend style={at={(0.98,0.97)}, anchor=north east, draw=black, fill=white, fill opacity=0.9, text opacity=1, font=\scriptsize, rounded corners=2pt},
  legend cell align=left,
  tick label style={font=\small},
  label style={font=\small},
  every axis plot/.append style={thick, mark size=2.2pt},
]
\addplot[color=closs, solid, mark=*, mark options={solid}]
  coordinates {(Tiny,0.943) (Small,0.256) (Medium,0.087) (Large,0.098)};
\addlegendentry{Loss}

\addplot[color=closs, dashed, mark=square*, mark options={solid}]
  coordinates {(Tiny,0.937) (Small,0.248) (Medium,0.080) (Large,0.103)};
\addlegendentry{Loss (EMA)}

\addplot[color=cmmd, solid, mark=*, mark options={solid}]
  coordinates {(Tiny,0.379) (Small,0.144) (Medium,0.071) (Large,0.076)};
\addlegendentry{MMD}

\addplot[color=cmmd, dashed, mark=square*, mark options={solid}]
  coordinates {(Tiny,0.378) (Small,0.138) (Medium,0.035) (Large,0.045)};
\addlegendentry{MMD (EMA)}

\end{axis}
\end{tikzpicture}
\caption{Validation loss and \ac{MMD} for \ac{FM} models of varying size, with and without \ac{EMA} of parameters. Solid lines: standard parameters; dashed lines: \ac{EMA} parameters. \ac{EMA} halves the \ac{MMD} at medium scale and reduces it by ${\approx}40\%$ for the large model, while having negligible effect on smaller models.}
\label{fig:fm_ema}
\end{figure}

\paragraph{Timestep sampling schedule.}
We compare five distributions for sampling the training timestep $t$: uniform $\mathcal{U}(0,1)$, log-normal (sigmoid of a standard normal), and three Beta distributions -- $\text{Beta}(2,1)$, $\text{Beta}(1,2)$, and $\text{Beta}(2,2)$ -- which, respectively, bias training toward $t \approx 1$ (near data), $t \approx 0$ (near noise), or the midpoint. All schedules yield nearly identical loss and \ac{MMD}, with differences below $0.005$ in both metrics. This insensitivity is consistent with a low-curvature velocity field: when the flow is close to a straight-line interpolation, all regions are roughly equally easy to fit and no schedule can substantially outperform uniform sampling. We retain the uniform schedule for simplicity.

\paragraph{Number of context observations.}
We evaluate the effect of providing $n \in \{1, 3, 5, 7\}$ random context observations to the \ac{FM} model. Both loss and \ac{MMD} increase monotonically with $n$, though the differences are small: loss grows from $0.075$ ($n=1$) to $0.080$ ($n=7$), while \ac{MMD} increases from $0.026$ to $0.035$. We attribute this to two factors: (i)~the context is generated by random transmissions, so additional observations are redundant given the first and primarily add noise to the conditioning signal, and (ii)~more context slots enlarge the input dimensionality per edge from $2d_z$ to $(n{+}1) d_z$, increasing the effective learning difficulty. We therefore set $n=1$ for all main experiments.

\paragraph{ODE solver and integration steps.}
Fig.~\ref{fig:fm_solver} compares the first-order Euler and second-order Heun solvers across the $1$--$20$ integration steps. Euler reaches \ac{MMD} of $0.033$ at 6~steps and converges to ${\approx}0.032$ beyond 10~steps, while Heun at 14~steps ($\text{\ac{MMD}}=0.034$) still does not match Euler's asymptote. Although counterintuitive given Heun's higher theoretical order, Euler consistently achieves lower \ac{MMD} at every step count tested. The results are consistent with a low-curvature velocity field, where first-order steps already trace the flow accurately; the Heun corrector term may introduce noise from querying the network slightly off the learned manifold. Since Euler also requires only one model evaluation per step (vs.\ two for Heun), it is faster and more accurate in this setting. We select the Euler solver with 6~steps for all experiments.

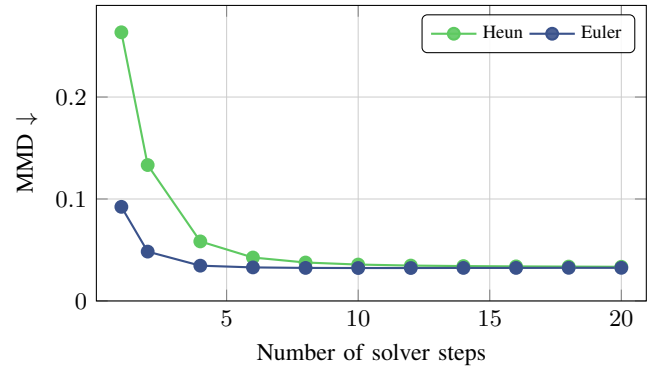
\begin{figure}[t]
\centering
\begin{tikzpicture}
\colorlet{cheun}{palD}
\colorlet{ceuler}{palA}

\begin{axis}[
  width=\columnwidth,
  height=0.62\columnwidth,
  xlabel={Number of solver steps},
  ylabel={MMD~$\downarrow$},
  ymin=0,
  grid=both,
  major grid style={line width=0.2pt, draw=gray!40},
  minor grid style={line width=0.1pt, draw=gray!20},
  enlarge x limits=0.05,
  clip=false,
  legend columns=2,
  legend style={at={(0.98,0.97)}, anchor=north east, draw=black, fill=white, fill opacity=0.9, text opacity=1, font=\scriptsize, rounded corners=2pt},
  legend cell align=left,
  tick label style={font=\small},
  label style={font=\small},
  every axis plot/.append style={thick, mark size=2.2pt},
]
\addplot[color=cheun, solid, mark=*, mark options={solid}]
  coordinates {
    (1,0.2635) (2,0.1333) (4,0.0584) (6,0.0426) (8,0.0377)
    (10,0.0357) (12,0.0347) (14,0.0342) (16,0.0339) (18,0.0337) (20,0.0335)
  };
\addlegendentry{Heun}

\addplot[color=ceuler, solid, mark=*, mark options={solid}]
  coordinates {
    (1,0.0923) (2,0.0484) (4,0.0346) (6,0.0329) (8,0.0324)
    (10,0.0323) (12,0.0323) (14,0.0324) (16,0.0324) (18,0.0325) (20,0.0325)
  };
\addlegendentry{Euler}

\end{axis}
\end{tikzpicture}
\caption{Comparison of Heun and Euler \ac{ODE} solvers for \ac{FM} generation as a function of integration steps. Euler achieves lower \ac{MMD} at every step count and converges in ${\approx}6$ steps, while Heun requires ${\approx}14$ steps without matching Euler's asymptote.}
\label{fig:fm_solver}
\end{figure}

\subsection{Surrogate Model}
\label{sec:surrogate_ablations}

\paragraph{Number of mixture components.}
Fig.~\ref{fig:surrogate_ablation_plot} shows the validation \ac{NLL} of the surrogate as a function of the number of components of the mixture~$m$. This ablation uses the non-embedded \ac{MDN} variant (graph input, without autoencoder pre-encoding). Across all model sizes, using mixtures ($m \geq 2$) substantially improves \ac{NLL} over a single Gaussian, and performance generally keeps improving with larger $m$: the best results are at $m=8$ for the tiny model and at $m=16$ for the small, medium, and large models. The medium model still achieves lower \ac{NLL} than the large model despite having fewer parameters, which suggests the large model may require a longer training budget to fully converge (Table~\ref{tab:optimizer_hparams}). The effective-component analysis (Fig.~\ref{fig:mdn_effective_components_boxplot}) shows increasing component utilization with larger mixtures: for $m=16$, the median effective component count is roughly 6.6--8.1 across model sizes. We retain $m=4$ in the main pipeline as a practical quality--complexity trade-off.

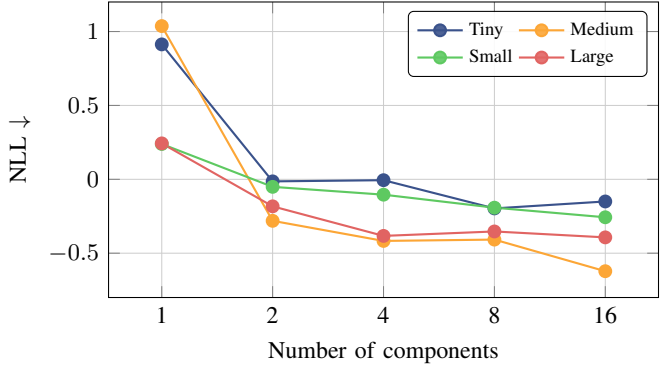
\begin{figure}[t]
\centering
\begin{tikzpicture}
\colorlet{ctiny}{palA}
\colorlet{csmall}{palD}
\colorlet{cmedium}{palH}
\colorlet{clarge}{palF}

\begin{axis}[
  width=\columnwidth,
  height=0.62\columnwidth,
  xlabel={Number of components},
  ylabel={NLL~$\downarrow$},
  ymin=-0.8, ymax=1.2,
  xmin=1, xmax=16,
  xtick={1,2,4,8,16},
  xticklabels={1,2,4,8,16},
  xmode=log,
  log basis x=2,
  grid=both,
  major grid style={line width=0.2pt, draw=gray!40},
  minor grid style={line width=0.1pt, draw=gray!20},
  enlarge x limits=0.12,
  clip=false,
  legend columns=2,
  legend style={at={(0.98,0.97)}, anchor=north east, draw=black, fill=white, fill opacity=0.9, text opacity=1, font=\scriptsize, rounded corners=2pt},
  legend cell align=left,
  tick label style={font=\small},
  label style={font=\small},
  every axis plot/.append style={thick, mark=*, mark size=2.2pt},
]
\addplot[color=ctiny] coordinates {(1,0.913) (2,-0.014) (4,-0.006) (8,-0.197) (16,-0.150)};
\addlegendentry{Tiny}

\addplot[color=cmedium] coordinates {(1,1.037) (2,-0.281) (4,-0.417) (8,-0.408) (16,-0.622)};
\addlegendentry{Medium}

\addplot[color=csmall] coordinates {(1,0.239) (2,-0.051) (4,-0.104) (8,-0.192) (16,-0.257)};
\addlegendentry{Small}

\addplot[color=clarge] coordinates {(1,0.243) (2,-0.183) (4,-0.383) (8,-0.353) (16,-0.393)};
\addlegendentry{Large}
\end{axis}
\end{tikzpicture}
\caption{Validation \ac{NLL} of the surrogate model as a function of the number of \ac{MDN} mixture components~$m$ for four model sizes. Note the logarithmic scale on the $X$ axis.}
\label{fig:surrogate_ablation_plot}
\end{figure}

\begin{figure}[t]
\centering
\begin{tikzpicture}
\colorlet{ctiny}{palA}
\colorlet{csmall}{palD}
\colorlet{cmedium}{palH}
\colorlet{clarge}{palF}

\begin{axis}[
  width=\columnwidth,
  height=0.62\columnwidth,
  ylabel={Effective no. of components},
  xlabel={Number of components},
  ymode=log,
  ytick={1,2,4,8,16},
  yticklabels={1,2,4,8,16},
  ymax=16,
  grid=both,
  major grid style={line width=0.2pt, draw=gray!40},
  minor grid style={line width=0.1pt, draw=gray!20},
  tick label style={font=\small},
  label style={font=\small},
  xtick={1,2.4,3.8,5.2,6.6},
  xticklabels={1,2,4,8,16},
  x coord trafo/.code={\pgfmathparse{1 + 0.7*(#1-1)}\pgfmathresult},
  x coord inv trafo/.code={\pgfmathparse{1 + (#1-1)/0.7}\pgfmathresult},
  x tick label style={yshift=-2pt},
  enlarge x limits=0.04,
  boxplot/draw direction=y,
  boxplot/variable width,
  boxplot/box extend=0.13,
  boxplot/every box/.style={solid},
  boxplot/every whisker/.style={solid},
  boxplot/every median/.style={solid},
  boxplot/every outlier/.style={solid, mark=*, mark size=1.8pt},
  legend style={at={(0.01,0.97)}, anchor=north west, draw=black, fill=white, fill opacity=0.9, text opacity=1, font=\scriptsize, rounded corners=2pt},
  legend cell align=left,
]


\addlegendimage{thick, mark=*, mark size=2.2pt, color=ctiny}
\addlegendentry{Tiny}
\addlegendimage{thick, mark=*, mark size=2.2pt, color=csmall}
\addlegendentry{Small}
\addlegendimage{thick, mark=*, mark size=2.2pt, color=cmedium}
\addlegendentry{Medium}
\addlegendimage{thick, mark=*, mark size=2.2pt, color=clarge}
\addlegendentry{Large}

%
%
%

\addplot+[
  boxplot prepared={draw position=2.08, median=1.278, lower quartile=1.086, upper quartile=1.726, lower whisker=1.006, upper whisker=2.000},
  draw=ctiny, fill=ctiny, fill opacity=0.18, thick, mark=*,
] coordinates {};

\addplot+[
  boxplot prepared={draw position=2.28, median=1.241, lower quartile=1.080, upper quartile=1.611, lower whisker=1.010, upper whisker=2.000},
  draw=csmall, fill=csmall, fill opacity=0.18, thick, mark=*,
] coordinates {};

\addplot+[
  boxplot prepared={draw position=2.52, median=1.346, lower quartile=1.201, upper quartile=1.593, lower whisker=1.022, upper whisker=2.000},
  draw=cmedium, fill=cmedium, fill opacity=0.18, thick, mark=*,
] coordinates {};

\addplot+[
  boxplot prepared={draw position=2.72, median=1.854, lower quartile=1.527, upper quartile=1.976, lower whisker=1.134, upper whisker=2.000},
  draw=clarge, fill=clarge, fill opacity=0.18, thick, mark=*,
] coordinates {};

\addplot+[
  boxplot prepared={draw position=3.48, median=2.974, lower quartile=2.548, upper quartile=3.453, lower whisker=1.621, upper whisker=4.000},
  draw=ctiny, fill=ctiny, fill opacity=0.18, thick, mark=*,
] coordinates {};

\addplot+[
  boxplot prepared={draw position=3.68, median=1.556, lower quartile=1.281, upper quartile=2.146, lower whisker=1.091, upper whisker=3.334},
  draw=csmall, fill=csmall, fill opacity=0.18, thick, mark=*,
] coordinates {};

\addplot+[
  boxplot prepared={draw position=3.92, median=2.315, lower quartile=1.805, upper quartile=2.672, lower whisker=1.282, upper whisker=3.507},
  draw=cmedium, fill=cmedium, fill opacity=0.18, thick, mark=*,
] coordinates {};

\addplot+[
  boxplot prepared={draw position=4.12, median=2.651, lower quartile=2.177, upper quartile=3.147, lower whisker=1.354, upper whisker=3.986},
  draw=clarge, fill=clarge, fill opacity=0.18, thick, mark=*,
] coordinates {};

\addplot+[
  boxplot prepared={draw position=4.88, median=4.787, lower quartile=4.214, upper quartile=5.189, lower whisker=2.752, upper whisker=6.413},
  draw=ctiny, fill=ctiny, fill opacity=0.18, thick, mark=*,
] coordinates {};

\addplot+[
  boxplot prepared={draw position=5.08, median=3.089, lower quartile=2.207, upper quartile=3.857, lower whisker=1.251, upper whisker=4.957},
  draw=csmall, fill=csmall, fill opacity=0.18, thick, mark=*,
] coordinates {};

\addplot+[
  boxplot prepared={draw position=5.32, median=2.896, lower quartile=2.325, upper quartile=3.417, lower whisker=1.627, upper whisker=4.926},
  draw=cmedium, fill=cmedium, fill opacity=0.18, thick, mark=*,
] coordinates {};

\addplot+[
  boxplot prepared={draw position=5.52, median=4.545, lower quartile=3.819, upper quartile=5.400, lower whisker=2.813, upper whisker=6.897},
  draw=clarge, fill=clarge, fill opacity=0.18, thick, mark=*,
] coordinates {};

\addplot+[
  boxplot prepared={draw position=6.28, median=8.074, lower quartile=6.141, upper quartile=8.832, lower whisker=3.200, upper whisker=9.849},
  draw=ctiny, fill=ctiny, fill opacity=0.18, thick, mark=*,
] coordinates {};

\addplot+[
  boxplot prepared={draw position=6.48, median=6.597, lower quartile=5.206, upper quartile=7.578, lower whisker=2.194, upper whisker=8.762},
  draw=csmall, fill=csmall, fill opacity=0.18, thick, mark=*,
] coordinates {};

\addplot+[
  boxplot prepared={draw position=6.72, median=6.930, lower quartile=4.005, upper quartile=8.012, lower whisker=1.751, upper whisker=9.935},
  draw=cmedium, fill=cmedium, fill opacity=0.18, thick, mark=*,
] coordinates {};

\addplot+[
  boxplot prepared={draw position=6.92, median=7.684, lower quartile=5.704, upper quartile=9.746, lower whisker=2.867, upper whisker=12.935},
  draw=clarge, fill=clarge, fill opacity=0.18, thick, mark=*,
] coordinates {};

\end{axis}
\end{tikzpicture}
\caption{Distribution of the effective number of active mixture components (entropy-based) for varying $m$ and model size. Note the logarithmic scale on both axes.}
\label{fig:mdn_effective_components_boxplot}
\end{figure}
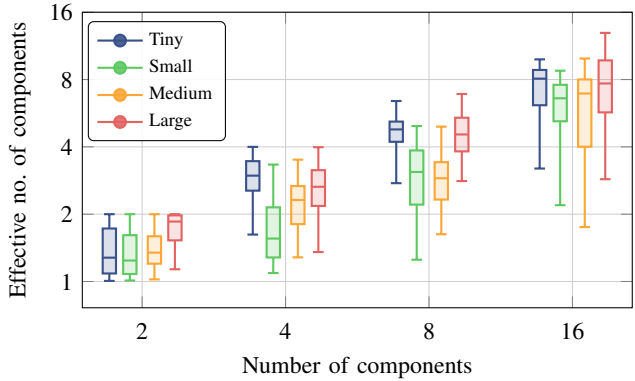

\paragraph{MDN vs. MSE prediction.}
Table~\ref{tab:surrogate_quality} reports the prediction quality of the two \ac{MDN} variants. The best result is $R^2=0.600$ (large, non-embedded \ac{MDN}), with correlations up to $0.774$ and biases in a relatively narrow range (\SIrange{-9.7}{+8.4}{\mega\bit\per\second}).
We also evaluate \ac{MSE}-trained point-estimate baselines (with and without autoencoder pre-encoding). They strongly overfit the training data and produce unreliable predictions on out-of-distribution inputs encountered during inference. We therefore adopt the \ac{MDN} formulation, which provides calibrated uncertainty estimates and more robust generalization.

\begin{table*}[t]
\caption{Surrogate prediction quality on the validation set. \ac{MDN} uses $m{=}4$ mixture components on non-embedded graph input, and \ac{MDN}+\ac{AE} applies \ac{GNN} autoencoder pre-encoding before the \ac{MDN} head. P90/P95/P99 denote absolute prediction error at the respective percentiles.}
\label{tab:surrogate_quality}
\centering
\small
\begin{tabular}{cc|cccc|ccc}
\toprule
Size & Type & $R^2$~$\uparrow$ & MAE [\si{\mega\bit\per\second}]~$\downarrow$ & Corr.~$\uparrow$ & Bias [\si{\mega\bit\per\second}] & P90 [\si{\mega\bit\per\second}]~$\downarrow$ & P95 [\si{\mega\bit\per\second}]~$\downarrow$ & P99 [\si{\mega\bit\per\second}]~$\downarrow$ \\
\midrule
\multirow{2}{*}{Tiny}
  & \ac{MDN}+\ac{AE}   & \textbf{0.450} & \textbf{177.9} & \textbf{0.674} & $-9.7$ & \textbf{408.0} & \textbf{499.2} & \textbf{646.8} \\
  & \ac{MDN}         & 0.331 & 201.0 & 0.582 & $\mathbf{-8.0}$ & 442.6 & 538.2 & 697.0 \\
\specialrule{0.4pt}{2pt}{2pt}
\multirow{2}{*}{Small}
  & \ac{MDN}+\ac{AE}   & \textbf{0.528} & \textbf{157.9} & \textbf{0.727} & $\mathbf{-0.1}$ & \textbf{371.9} & \textbf{466.0} & 674.2 \\
  & \ac{MDN}         & 0.485 & 169.9 & 0.698 & $-5.0$ & 393.1 & 480.9 & \textbf{643.6} \\
\specialrule{0.4pt}{2pt}{2pt}
\multirow{2}{*}{Medium}
  & \ac{MDN}+\ac{AE}   & \textbf{0.551} & \textbf{150.2} & \textbf{0.743} & $+8.4$ & \textbf{365.6} & \textbf{453.5} & 684.8 \\
  & \ac{MDN}         & \textbf{0.552} & 152.2 & \textbf{0.743} & $\mathbf{-2.4}$ & 367.0 & 460.0 & \textbf{638.7} \\
\specialrule{0.4pt}{2pt}{2pt}
\multirow{2}{*}{Large}
  & \ac{MDN}+\ac{AE}   & 0.593 & 142.9 & 0.771 & $+7.6$ & \textbf{349.2} & 439.2 & 617.2 \\
  & \ac{MDN}         & \textbf{0.600} & \textbf{139.9} & \textbf{0.774} & $\mathbf{+3.0}$ & \textbf{349.4} & \textbf{437.8} & \textbf{609.3} \\
\bottomrule
\end{tabular}
\end{table*}

\paragraph{Autoencoder pre-encoding.}
For the \ac{MDN} variants, pre-encoding improves tiny/small ($R^2$: $0.331 \rightarrow 0.450$, $0.485 \rightarrow 0.528$) but is neutral to slightly negative at medium/large ($0.552 \rightarrow 0.551$, $0.600 \rightarrow 0.593$). Overall, latent pre-encoding helps low-to-mid capacity models more consistently than the largest ones. Since the main pipeline already uses the frozen \ac{GNN} encoder for the \ac{FM} model, we retain pre-encoding for the surrogate (\ac{MDN}+\ac{AE}) as it adds no extra cost and benefits smaller models.

\paragraph{Scaling and tail errors.}
Taking the best \ac{MDN} model per size, $R^2$ rises from $0.450$ (tiny) to $0.600$ (large), indicating consistent improvement with model capacity. Table~\ref{tab:surrogate_quality} also reports absolute-error tails at the 90th, 95th, and 99th percentiles. Tail errors decrease consistently with model size: the P90 error drops from \SI{408}{\mega\bit\per\second} (tiny) to \SI{349}{\mega\bit\per\second} (large), while the P99 error remains in the \SIrange{609}{697}{\mega\bit\per\second} range across all configurations. The two \ac{MDN} variants exhibit similar tail behavior at medium and large scales, whereas \ac{MDN}+\ac{AE} has noticeably tighter tails at tiny and small sizes, consistent with the pre-encoding benefit observed in point-prediction metrics.

\paragraph{Risk-averse candidate selection.}
Since the \ac{MDN} surrogate provides predictive uncertainty via the mixture variance $\bar{\sigma}_e^2 = \sum_k w_k (\sigma_k^2 + \mu_k^2) - \bar{r}_e^2$, we test whether penalizing uncertain predictions improves candidate selection. We score candidates by $s = \bar{r}_e - \lambda \cdot \bar{\sigma}_e$, where $\lambda \geq 0$ controls the degree of risk aversion, and evaluate the effect on the effective data rate (Fig.~\ref{fig:risk_factor}). Risk-neutral selection ($\lambda=0$) yields \SI{1136.4}{\mega\bit\per\second}, while the best risk-averse setting ($\lambda=0.5$) achieves \SI{1139.1}{\mega\bit\per\second} -- an improvement of only \SI{0.24}{\percent}. For $\lambda \geq 1.5$, performance drops sharply as the criterion becomes overly conservative. Given the negligible benefit, we use risk-neutral selection ($\lambda=0$) in all main experiments.

\begin{figure}[t]
\centering
\begin{tikzpicture}
\colorlet{cdarkblue}{palA}

\begin{axis}[
  width=\columnwidth,
  height=0.62\columnwidth,
  xlabel={$\lambda$},
  xmin=0, xmax=4,
  ylabel={Effective data rate [\si{\mega\bit\per\second}]~$\uparrow$},
  xtick={0,0.5,1.0,1.5,2.0,2.5,3.0,3.5,4.0},
  ymin=0, ymax=1500,
  ytick={0,300,600,900,1200,1500},
  grid=both,
  major grid style={line width=0.2pt, draw=gray!40},
  minor grid style={line width=0.1pt, draw=gray!20},
  enlarge x limits=0.08,
  clip=false,
  tick label style={font=\small},
  label style={font=\small},
  every axis plot/.append style={thick, mark size=2.2pt},
]
\addplot[color=cdarkblue, dashed, thick, forget plot]
  coordinates {(-0.3,1350.066) (4.3,1350.066)};
\node[anchor=south west, font=\scriptsize] at (axis cs:0.0,1320) {Optimal};

\addplot[color=cdarkblue, dotted, thick, forget plot]
  coordinates {(-0.3,479.469) (4.3,479.469)};
\node[anchor=north west, font=\scriptsize] at (axis cs:0.0,480) {Random};

\addplot[color=cdarkblue, solid, mark=*, mark options={solid}]
  coordinates {(0.0,1136.370) (0.5,1139.108) (1.0,1138.656) (1.5,1080.381) (2.0,941.683) (2.5,826.825) (3.0,704.537) (3.5,703.007) (4.0,702.712)};

\end{axis}
\end{tikzpicture}
\caption{Effect of the risk factor $\lambda$ on mean effective data rate. Dashed line: best configuration for the scenario (\SI{1350.1}{\mega\bit\per\second}); dotted line: random selection (\SI{479.5}{\mega\bit\per\second}).}
\label{fig:risk_factor}
\end{figure}
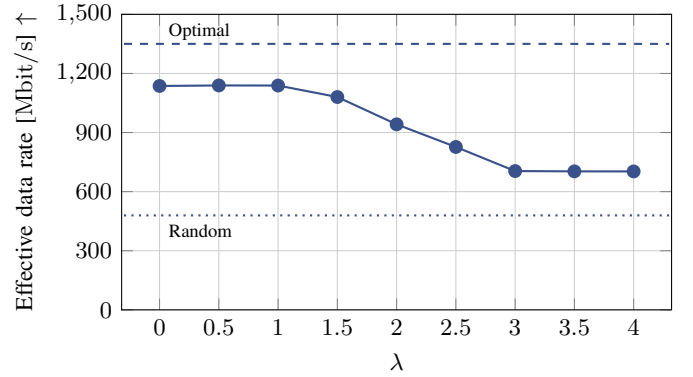


\section{Computational Overhead}

Table~\ref{tab:convergence_times} compares the time required to produce a \ac{Co-SR} schedule across methods and scales. FM4WiFi has no online convergence phase: all learning is performed offline, and the online cost is limited to preparing batched inputs and running the compiled inference pipeline (128 candidates in 4 sequential batches of 32, as discussed in Section~\ref{sec:scalability}). This results in sub-second latency at all tested scales, growing only from \SI{0.13}{\second} at 2x2 to \SI{0.62}{\second} at 4x4.

\begin{table}[t]
  \caption{Time to produce a \ac{Co-SR} schedule in the residential scenario (room size \SI{10}{\meter}). For FM4WiFi and analytical baselines, this is a one-shot inference time; for \ac{H-MAB}, it is the online convergence time (\textit{inf} = did not converge within the time budget).}
  \label{tab:convergence_times}
  \centering
  \small
  \begin{tabular}{c|ccc}
    \toprule
    Method           & 4 APs (2x2)        & 9 APs (3x3)         & 16 APs (4x4)       \\
    \midrule
    T-Optimal        & \SI{0.34}{\second} & \SI{2.58}{\second}  & \SI{29.7}{\second} \\
    F-Optimal        & \SI{1.24}{\second} & \SI{279.2}{\second} & \SI{2956}{\second} \\
    H-MAB            & \SI{3.78}{\second} & \SI{78.97}{\second} & \textit{inf}       \\
    FM4WiFi          & \SI{0.13}{\second} & \SI{0.24}{\second}  & \SI{0.62}{\second} \\
    \bottomrule
  \end{tabular}
\end{table}

In contrast, the other methods exhibit exponential scaling. T-Optimal reaches \SI{29.7}{\second} at 4x4 -- too slow for practical use. F-Optimal takes nearly 50~minutes on the same scale, and \ac{H-MAB} does not converge within any practical budget. FM4WiFi is therefore the only method that remains tractable on all evaluated scales, and its constant-time character with respect to topology changes makes it directly suitable for the event-triggered deployment demonstrated in Fig.~\ref{fig:dynamic_scenarios_enterprise}.

\section{Additional Results}
\label{sec:additional_results}

Figures~\ref{fig:dynamic_individual_once}--\ref{fig:dynamic_individual_hmab} show per-seed throughput traces for the wall-free residential scenario (Fig.~\ref{fig:dynamic_scenarios_res_no_walls} in the main text), decomposing the averaged curves into the 5 individual runs. All runs share the same topology and mobility pattern, so the observed variance reflects algorithmic stochasticity. FM4WiFi (with MCS selection) (Figures~\ref{fig:dynamic_individual_once} and~\ref{fig:dynamic_individual_interval_1s}) shows significant variance both between and within runs, as its probabilistic sampling produces different \ac{Co-SR} configurations each time. \ac{H-MAB} (Fig.~\ref{fig:dynamic_individual_hmab}) is the most stable across runs -- not because it converges well, but because it barely converges at all in this large scenario, maintaining a roughly constant throughput.

\begin{figure}[t]
\centering
\input{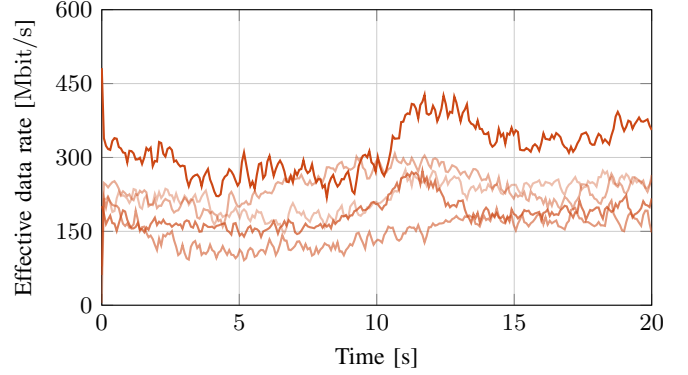}
\caption{Per-seed effective data rate for FM4WiFi without regeneration, 4x4 residential scenario (room size \SI{20}{\meter}, without walls).}
\label{fig:dynamic_individual_once}
\end{figure}

\begin{figure}[t]
\centering
\input{figures/dynamic_individual/dynamic_individual_interval_1.tikz}
\caption{Per-seed effective data rate for FM4WiFi with cyclic regeneration every \SI{1}{\second}, 4x4 residential scenario (room size \SI{20}{\meter}, without walls).}
\label{fig:dynamic_individual_interval_1s}
\end{figure}

\begin{figure}[t]
\centering
\input{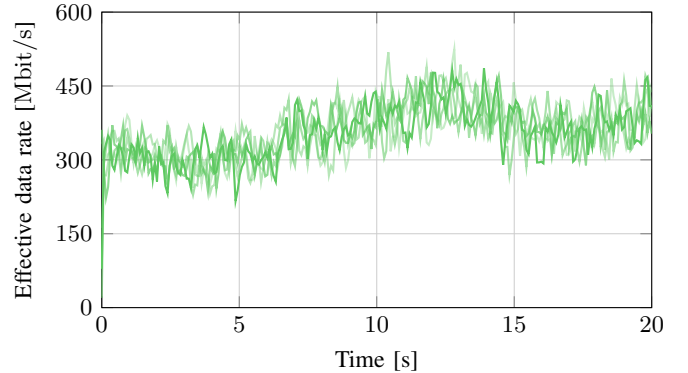}
\caption{Per-seed effective data rate for \ac{H-MAB}, 4x4 residential scenario (room size \SI{20}{\meter}, without walls).}
\label{fig:dynamic_individual_hmab}
\end{figure}

\fi 

\end{document}